\PassOptionsToPackage{square,comma,sort&compress,super, authoryear}{natbib}  

\documentclass[%
preprint,
nofootinbib,
bibnotes,
 amsmath,amssymb,
 aps,
floatfix,
]{revtex4-2}

\usepackage{subcaption}
\usepackage{layout}
\usepackage{amsmath}
\usepackage{comment}
\usepackage{float}
\usepackage{hyperref}
\usepackage{cleveref}
\usepackage{ccicons}
\newcommand{\Qc}{\dot Q_{\text{c}}}
\newcommand{\Qh}{\dot Q_{\text{h}}}

\newcommand{\Upv}{U_{\text{PV}}}
\newcommand{\Uled}{U_{\text{LED}}}
\newcommand{\Tpv}{T_{\text{PV}}}
\newcommand{\Tled}{T_{\text{LED}}}
\newcommand{\kB}{k_{\text{B}}}

\newcommand{\led}{_{\text{LED}}}
\newcommand{\pv}{_{\text{PV}}}

\newcommand{\s}[1]{_{\text{#1}}}

\newcommand{\Qabove}{\dot Q_{\text{rad}}^{>}}
\newcommand{\Qsub}{\dot Q_{\text{rad}}^{<}}
\newcommand{\Qtot}{\dot Q_{\text{rad,net}}}
\newcommand{\eqeled}{\mathrm{EQE}_{\mathrm{LED}}}
\newcommand{\iqeled}{\mathrm{IQE}_{\mathrm{LED}}}
\newcommand{\Eg}{E_\text{g}}
\newcommand{\CoolDens}{W.m$^{-2}$}

\newcommand{\Ppv}{P_{\text{PV}}}
\newcommand{\Pled}{P_{\text{LED}}}
\newcommand{\Pelec}{P_{\text{in}}}
\makeatletter
\newcommand\freefootnote[1]{%
	\begingroup
	\let\@thefnmark\relax
	\renewcommand\thefootnote{}%
	\footnotetext{#1}%
	\endgroup
}
\makeatother

\usepackage{graphicx}% Include figure files
\usepackage{dcolumn}% Align table columns on decimal point
\usepackage{bm}% bold math

\usepackage[table]{xcolor}

\definecolor{purple2}{RGB}{139,0,139} % RGB format
\definecolor{DarkGreen}{RGB}{0, 125, 0}

\graphicspath{./}

\begin{document}

\title{Performances of far and near-field thermophotonic refrigeration devices from the detailed-balance approach}
\thanks{CETHIL : CNRS - Insa Lyon - Université Claude Bernard Lyon 1}%

%\author{Thomas Châtelet, Julien Legendre, Olivier Merchiers, Pierre-Olivier Chapuis}
%\altaffiliation[Also at ]{CETHIL, Villeurbanne.}
%\affiliation{CETHIL - CNRS - INSA Lyon - Université Claude-Bernard Lyon 1}

\color{blue}
\author{Thomas Châtelet - Corresponding author*}
\email{thomas.chatelet@protonmail.com}
\affiliation{%
	CETHIL - CNRS - INSA Lyon - Université Claude-Bernard Lyon 1,\\
	9, rue de la Physique,\\ 
	France, 69621 Villeurbanne cedex\\ 
	thomas.chatelet@protonmail.com
}

\author{Julien Legendre}
\email{jlegendre@icfo.net}
\affiliation{
ICFO ,  3 Av. Carl Friedrich Gauss, \\ 
Spain, Barcelona 08860 Castelldefels
}

\author{Olivier Merchiers}
\email{olivier.merchiers@insa-lyon.fr}
\affiliation{CETHIL - CNRS - INSA Lyon \\
	9, rue de la Physique,\\ 
	France, 69621 Villeurbanne cedex
}

\author{Pierre-Olivier Chapuis}
\email{pierre-olivier.chapuis@insa-lyon.fr}
\affiliation{CETHIL - CNRS - INSA Lyon \\
	9, rue de la Physique,\\ 
	France, 69621 Villeurbanne cedex
}
\color{black}

\date{\today}

\begin{abstract}

We study a near-field thermophotonic (NF-TPX) refrigerating device, consisting of a light-emitting diode and a photovoltaic cell in close proximity. Calculations are performed in the frame of the detailed-balance approach. We study how thermal radiation, separation distance and LED temperature can affect both cooling power and coefficient of performance. More specifically, we assess the impact of bandgap energy and quantum efficiency for an artificial material on those cooling performances. For a particular device made of GaAs and/or AlGaAs we show that, in the near-field regime, the cooling power can be increased by one order of magnitude compared to far field. However, a 10\% reduction of the quantum efficiency can lead to a decrease of the cooling power by two orders of magnitude. Finally, we compare existing literature data on electroluminescent, TPX and thermoelectric cooling with our detailed balance prediction, which highlights design-rule requirements for NF-TPX cooling devices.
\end{abstract}

\maketitle

%\tableofcontents
\section{Introduction}
\freefootnote{Creative Commons Attribution 4.0 International License. \href{https://creativecommons.org/licenses/by/4.0/}{\ccby} }
Cooling technologies have a wide range of applications including ventilation, air conditioning, household refrigerators, electronic component thermal management and cryogenic control of nanoscale devices down to atomic systems. The expected rise of quantum technologies will require high-quality cooling devices capable of operating in diverse conditions and adapted for micro- and nanoscale integration. Currently, the most widely used cooling systems rely on vapour compression and are very well suited for large systems such as building interiors, cooling chambers or refrigerators. At smaller scale, however, vapour compression suffers from drawbacks due to noise and vibrations caused by piston or rotating compressors, which furthermore require maintenance. Solid-state cooling devices, such as thermoelectric devices (TEC), avoid those drawbacks and are more easily integrated within small-scale devices. They are one possible path to micro- and nanoscale cooling \cite{Goldsmid2009}. These, however, suffer from low coefficients of performances (COP) for temperature differences larger than 10 K \cite{Mao2021,Sun2022} and require a certain thickness in order to maintain the temperature difference between hot and cold sides.

In an attempt to solve those issues, a new class of solid-state devices based on photonic cooling has been proposed \cite{Duraffourg1965,Sheik-Bahae2007,Imangholi2007,Seletskiy2016}. Some of these systems are referred to as electroluminescent cooling (ELC) systems, relying on the use of light-emitting diodes (LED) to achieve cooling \cite{Dousmanis1964,Berdahl1985, Tauc1957,Park2024}. The electroluminescent cooling regime leads, under certain conditions, to heat extraction from the environment, usually the LED's crystalline lattice. One way of improving this system is through an LED combined with photovoltaic cell (PV) separated by a vacuum gap. The PV cell collects the radiation emitted by the LED and converts it to electricity, which can be fed back to the LED and thereby reduce the required external power, improving the coefficient of performance (COP) of the whole system \cite{Sadi2020}. This combination is known as thermophotonic (TPX) \cite{Harder2003} system (see Fig. \ref{fig:Systempic}). Here, heat exchange occurs only through radiative transfer, so it is expected to sustain larger temperature differences \cite{Sadi2020} than Peltier modules for instance. In this work we analyze the potential of near-field radiative transfer to improve cooling performances. Near field is known to increase by orders of magnitude the radiative transfer between two objects separated by distances smaller than Wien's wavelength in comparison to the transfer predicted by Stefan-Boltzmann's law. This enhancement comes from evanescent modes present only at the objects surface \cite{Mulet2002,Joulain2005,Greffet2007,Reid2013}. It was shown that near field effects can also increase electroluminescent radiation transfer, either for energy harvesting \cite{Zhao2018,Legendre2022,Legendre2022PIN} or for refrigeration \cite{Liu2016,Liao2022,Song2020,Li2021,Yang2022,PatrickXiao2018,Liao2019,Lin2018,Zhou2020,Zhu2019}. 
The cited works referring to electroluminescent cooling use planar configurations. Electrical transport is computed from the detailed-balance limit and radiative transfer from the fluctuational electrodynamics formalism \cite{Chen2017,Park2024}. Cooling performances vary strongly with gap distance, temperature difference in the system, the geometry involving multilayers and the nature of the emitter-receiver pair. However, little research on the TPX system as a cooling device has been conducted compared to electroluminescent cooling. 
The aim of the present work is to estimate the maximum theoretical limits of the cooling power and COP for an idealized TPX cooling system using the detailed-balance approach, and to understand how the device quality and near-field transfer impact performances. The impact of using a PV cell as a second heat engine in the system will be addressed, especially in light of the necessary trade-off between cooling power and COP. In addition, the effect of distance and temperature difference on the cooling power conditions will be investigated. \\
 The manuscript is structured as follows: we begin by a description of the energy balance in the system, we then provide a theoretical overview of the detailed-balance approach applied to a near-field TPX system and end this section by giving the relevant figures of merit. In the second part we analyse the results and start with the study case of an artificial material to understand the effect of bandgap energy and quantum efficiency (QE). We then focus on GaAs-based devices, studying the cooling power and COP as a function of vacuum gap size, LED temperature and QE. We compare our results with the state of the art and perform a critical analysis of the results to extract design rules for TPX cooling systems, suggesting few prospects.

\section{Idealized thermophotonic system}
\begin{figure}[t]
	\centering
	\includegraphics[width=400pt]{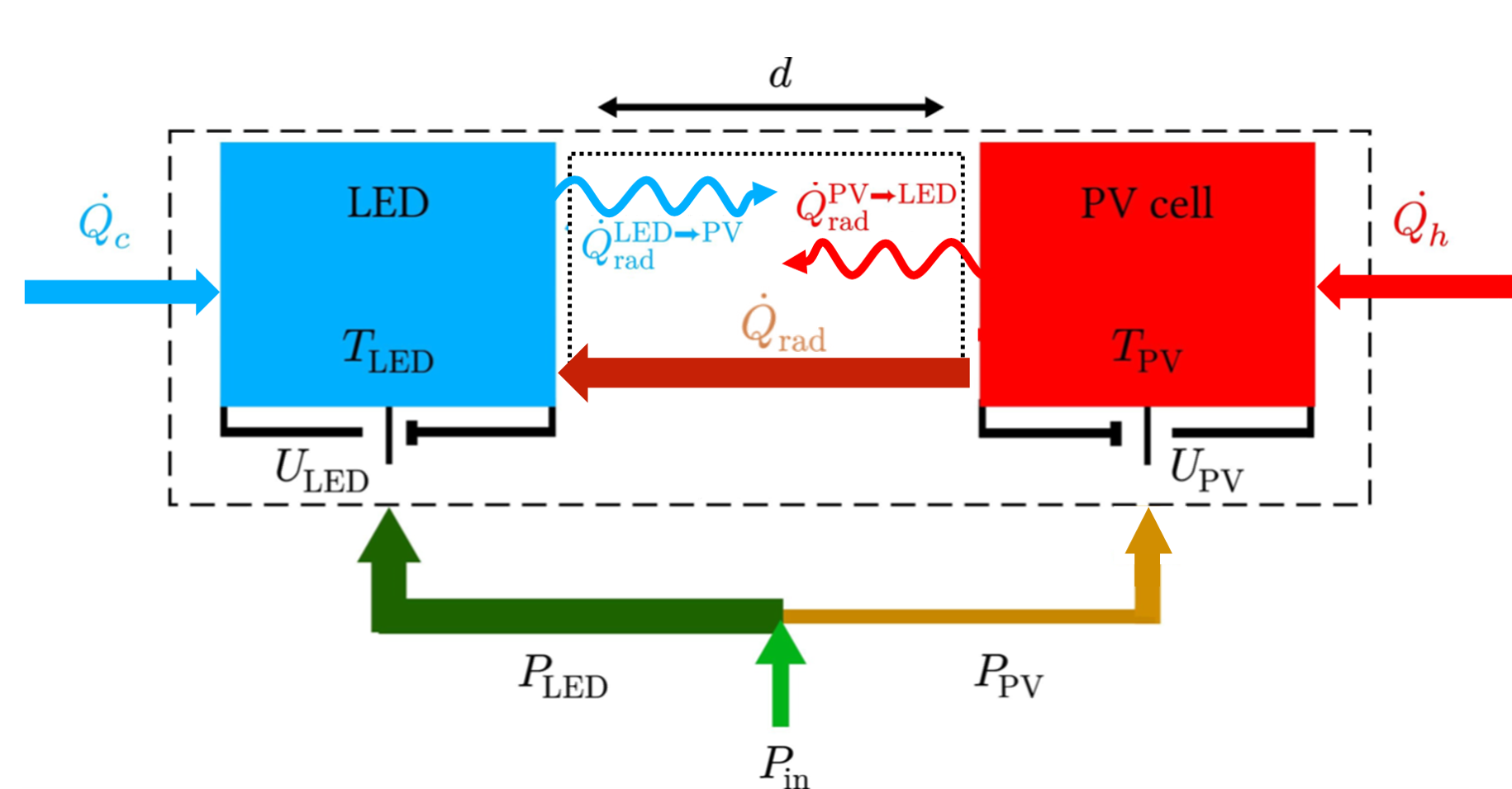}
	\caption{Schematic of the TPX cooling system. The energy is taken positive when entering the system and negative when leaving.}
	\label{fig:Systempic}
\end{figure}

We consider an idealized system in which the LED and the PV cell are homogeneous planar semi-infinite media separated by a vacuum gap of size $d$ (Fig. \ref{fig:Systempic}). The detailed-balance approach does not require electrical properties of the materials, except for the bandgap energy which is set as a parameter or taken from the literature when considering specific materials. Similarly, the QE, which is defined as the ratio of radiative recombinations (electron-hole pair generating a photon) to the total density of recombinations (including those that do not lead to light emission), is a parameter that is not computed directly from recombination mechanisms for both the LED and the PV cell. Throughout this work we set PV cell temperature to $\Tpv = 300$ K and for the LED we consider $\Tled < 300$ K.

To estimate the cooling power, we consider the energy balance at steady state. Energy fluxes are shown in Fig. \ref{fig:Systempic}. For the whole TPX device we can write
\begin{equation}
	\Qc + \Pelec + \Qh = 0,
	\label{eq:Global_balance}
\end{equation}
where $\Qc$ is the cooling power extracted from the cold side, $\Qh$ is the heat flux dissipated to the room-temperature heat sink and $\Pelec$ the electrical power fed to the whole system. Considering only energy conservation for the LED, gives  
\begin{equation}
	\Qc + P\led + \dot  Q\s{rad,net} = 0,
	\label{eq:LED_balance}
\end{equation}
where $ \dot Q\s{rad,net}$ is the net radiative heat flux exchanged between LED emitter and PV cell receiver, $P\led$ is the electrical power fed to the LED and $\Qc$ is the cooling power. For clarity, $\Pled > 0$ means that electrical power is fed into the LED. A specific case is $\Qtot < 0$, which means that radiative heat flux goes from the LED to the PV cell. Cooling of the LED occurs when $\Qc > 0$ and as a consequence $P\led < - \dot Q\s{rad,net}$ indicating that the LED emits more radiative energy than it receives electrical energy. This is the so-called electroluminescent cooling regime. Finally,
\begin{equation}
	\Pelec=P\led + P\pv = U\led J\led + U\pv J\pv,
\end{equation}
where $J\led$ and $J\pv$ stand for the current densities within each device. $U\led$ and $U\pv$ are the LED and PV cell biases respectively. The performance characteristics are described by the cooling power $\Qc$ and the coefficient of performance (COP). The latter is given by 
\begin{equation} 
	\text{COP} = \frac{\Qc}{\Pelec}.
	\label{eq:COP}
\end{equation}
Since the LED emits in the electroluminescent regime, $P\led > 0$. For the PV cell, power is harvested, hence $P\pv < 0$ which reduces $\Pelec$ in absolute value leading to an improved COP in comparison to electroluminescent cooling. We also consider the scaled COP (SCOP):
\begin{equation} 
	\text{SCOP} = \frac{\text{COP}}{\mathrm{COP}\s{carnot}}, 
	\label{eq:SCOP}
\end{equation}
where $	\text{COP}_{\text{carnot}}$ is the upper bound given by
\begin{equation}
	\text{COP}_{\text{carnot}} = \frac{T\led}{T\pv-T\led}.
	\label{eq:epsCarnot}
\end{equation}

\section{Detailed-balance approach and near-field radiative transfer}

\subsection{Radiative flux calculations}
Since we are interested in near-field effects, we use the Fluctuational Electrodynamics (FE) framework to compute all contributions (propagative and evanescent) rigorously.
The total radiative heat flux leaving the LED is expressed as 
\begin{equation}
     \Qtot = \Qabove + \Qsub.
	\label{eq:Qtot}
\end{equation}
$\Qabove$ and $\Qsub$ are the net exchanged flux density above and below the energy bandgap of the considered materials. We consider the case in which the bandgap of the emitter and the receiver are matched. The fluxes can be expressed as
\begin{align}
	\Qsub &= \int_{0}^{\omega\s{gap}} \left[\Theta(T\led,\omega)-\Theta(T\pv,\omega)\right]\tau\s{tot} (\omega)  \:d\omega, 
	\label{eq:Qsub} \\
	\Qabove &= \int_{\omega\s{gap}}^{+\infty} \left[\Theta(T\led,U\led,\omega)-\Theta(T\pv,U\pv,\omega)\right]\tau\s{tot}(\omega)  \:d\omega, 
	\label{eq:Qabove}
\end{align}
where $\Theta$ is the mean energy of the generalized Planck oscillator given by \cite{Wurfel1982}
\begin{equation}
\Theta(T,U,\omega) = \begin{cases}
	\frac{\hbar\omega}{\exp{\left(\frac{\hbar\omega}{\kB T}\right)} - 1} & \text{if $\hbar\omega < E\s{g}$},\\
	\frac{\hbar\omega}{\exp{\left(\frac{\hbar\omega - eU}{\kB T}\right)} - 1} & \text{if $\hbar\omega \geq  E\s{g}$},
\end{cases}
\end{equation}
where $e$ is the elementary charge and $E\s{g}$ the bandgap energy. The above-bandgap  radiation involves electroluminescence while the sub-bandgap radiation is due to thermal radiation only.
The transfer coefficient $\tau\s{tot}(\omega)$ includes all information on geometry and optical properties of the media. In our case, we consider only planar homogeneous media parallel with one another, for which it is possible to compute the transfer coefficient using a semi-analytical method. We implemented the S-matrix method as described in \cite{Francoeur2009}, which allows to describe multilayer stacks and to calculate the exchanged flux with a single layer of the stack taking into account the near-field effects. For planar media, the transfer coefficient is the sum of far- and near-field contributions:
\begin{equation}
	\tau\s{tot}(\omega) = \tau\s{far-field}(\omega) + \tau\s{near-field}(\omega),
\end{equation} 
where 
\begin{align}
	\tau\s{far-field}(\omega) &= \int_0^{\omega/c} \mathcal{T}\s{ff}(\omega, k_{\rho}) k_{\rho} dk_{\rho}, \\
	\tau\s{near-field}(\omega) &= \int_{\omega/c}^{\infty} \mathcal{T}\s{nf}(\omega, k_{\rho}) k_{\rho} dk_{\rho}.
\end{align}
$k_{\rho}$ is the component of the wavevector parallel to the interfaces and $\mathcal{T}\s{ff,nf}$ the monochromatic and directional transmission coefficient.

%Here, the bandgap energy of the material at is approximately $1.42$ eV corresponding with $\omega_{gap}  \approx  2.16  \:×  \:10^{15}  \: rad.s^{-1}  \:$.  

\subsection{Current densities and electrical power calculations}

We compute $\Pled$ and $\Ppv$ in the frame of the detailed-balance approach which relates to the generation and recombination of electron-hole pairs in the semiconductors, and respectively absorption or emission of radiation.   
We consider first the radiative limit, in which only radiative recombinations occur (QE = 1). For the PV cell, this means that for each absorbed photon with an energy equal or higher than the bandgap, exactly one electron-hole pair is created and transferred to the load. For the LED, this means that every supplied electron-hole pair recombines by emitting a photon. To obtain the electrical power, we need to compute first the current density fed to the LED and extracted from the PV cell. The current density can be written as \cite{Legendre2022} 
\begin{equation}
	\centering
	J = \int_{\omega\s{gap}}^{+\infty}  e  \: \gamma\s{net}(\omega)  \:d\omega,
	\label{eq:Jperfect}
\end{equation}
where the net spectral photon flux density $\gamma\s{net}(\omega)$ in the radiative limit is expressed as 
\begin{equation}
	\centering
	\gamma\s{net}(\omega) = \left[ \Theta(T_{\mathrm{LED}},U_{\mathrm{LED}},\omega)-\Theta(T_{\mathrm{PV}},U_{\mathrm{PV}},\omega)\right] \dfrac{\tau\s{tot}(\omega)}{\hbar\omega}  \:.
	\label{eq:photonperfect}
\end{equation}

\subsection{Inclusion of nonidealities in detailed-balance approach}

In real-world applications, several nonideal factors affect the system performance. Not all charge carriers in the LED recombine to produce photons, not all photons reaching the PV cell are absorbed and generated into electron-hole pairs.
 In addition, not all generated electron-hole pairs contribute to the reduction of the external electrical power ($\Pelec$) required for system operation. As a result, we include nonidealities by means of \color{black}the external quantum efficiency for LEDs (different definition than that for PV cells) $\eqeled$\color{black}, which quantifies how efficiently charge carriers are converted to photons in the LED and how effectively photons are converted back to usable electron-hole pairs in the PV cell. While \color{black}$\eqeled$\color{black} is primarily determined by the intrinsic properties of the materials used, it can also be influenced by the manufacturing process and device structure. Improvements in these areas can enhance the overall efficiency of the system.
\color{black}$\eqeled$\color{black} can be simply expressed as
\begin{equation}
	\centering
	\mathrm{\color{black}\eqeled\color{black}} = \dfrac{n\s{r}}{n\s{r}+n\s{nr}},
\end{equation}
%$QE = \dfrac{number of radiative recombinations}{All Recombinations}$
where $n_{\text{r}}$ is the density per unit of time of radiative recombinations \color{black} that escapes one optoelectronic device \color{black} and $n_{\text{nr}}$ is the density per unit of time of nonradiative recombinations in a given device. The last quantity stands for an electron-hole pair recombination that does not create a photon. In the radiative limit we have $n\s{nr} = 0$. \color{black} Note that $\eqeled$ is also termed external radiative efficiency (ERE) in photovoltaics \cite{Rau2007,Steiner2013,DeSutter2017,Aeberhard2021}. Such formulation involves a two-point Green tensor that describes both the emitting and receiving positions, and therefore corresponds to external luminescence. If \color{black}$\eqeled$\color{black} $ \, <$ 1, some charge carriers contribute to heating in the LED. By inverting the equation, we obtain
\begin{equation}
	\centering
	n\s{nr} = \left( \dfrac{1}{\mathrm{\color{black}\eqeled\color{black}}} - 1 \right) n\s{r}.
\end{equation}
The current density $J\led$ is defined as
\begin{equation}
	\centering
	J\led = e\left( g - n \right),
	\label{BilanQE}
\end{equation}
in which $g = g\s{r} + g\s{nr} $ is the total density of electron-hole generation (due to radiative and nonradiative events) and $n = n\s{r} + n\s{nr}$ is the total number density of nonradiative and radiative electron-hole recombinations. When no voltage is applied (equilibrium case), nonradiative generation should cancel nonradiative recombination \cite{Harder2003}, so 
\begin{align}
g\s{nr} &= n\s{nr} (U = 0),  \\
g\s{nr} &= \left( \dfrac{1}{\mathrm{\color{black}\eqeled\color{black}}} - 1 \right) n\s{r}(U = 0).
\end{align}
\color{black}Using this expression\color{black}, one gets
\begin{equation}
	\centering
	J\led = e\left( g\s{r} - n\s{r} - \left( \dfrac{1}{\mathrm{\color{black}\eqeled\color{black}}} - 1\right)  \left( n\s{r} - n\s{r}(U = 0)\right) \right).
	\label{BilanQE}
\end{equation}
This expression is valid spectrally. As $ \gamma\s{net}(\omega) = g\s{r}(\omega) - n\s{r}(\omega)$ \color{black}and considering photons escaping the LED \color{black} $n\s{r}(\omega) = \Theta(\Tled,\Uled,\omega) \dfrac{\tau\s{tot}(\omega)}{\hbar\omega} $, we can obtain the current density, which is lower than that in the radiative limit (Eq. \ref{eq:Jperfect}) \cite{Legendre2022,Sadi2020,Harder2003}: 
\begin{equation}
	\centering
	J\led = \int_{\omega\s{gap}}^{+\infty}  e  \: \left[ \gamma\s{net}(\omega) - \left( \color{black}\dfrac{1}{\mathrm{\eqeled}}\color{black} - 1\right)  \left[ \Theta(\Tled,\Uled,\omega)-\Theta(\Tled,0,\omega)\right]  \dfrac{\tau\s{tot}(\omega)}{\hbar\omega} \right]  \:d\omega .
	\label{eq:Jnonperfect}
\end{equation}
\color{black} By contrast, internal quantum efficiency (IQE) refers to the radiative recombination fraction within the emitting material itself. The relationship between $\iqeled$ and $\eqeled$ can be written as $\eqeled = \text{LEE} \times \iqeled$, where LEE is the light extraction efficiency. LEE quantifies the proportion of photons that exits an optoelectric device amongst generated photons inside a device. It must be reminded that achieving $\eqeled=1$ in practice is extremely challenging, especially in the far field, due to unavoidable nonradiative losses and imperfect light extraction in III-V materials \cite{Giannini2021}. In the near-field regime, however, the strong increase of radiative exchange compared to nonradiative processes brings this limit closer to reality \cite{Papadakis2021}. This distinction underlines the practical challenge of realizing far-field thermophotonic devices, as also suggested by recent drift–diffusion modeling of harvesting configurations \cite{Legendre2022,legen2023}.
\color{black}
In our work, we consider a symmetric situation i.e. that the same \color{black}$\eqeled$ $\, $\color{black} is applied in the LED and the PV cell. Note that, in literature, \color{black}$\eqeled $ $\,$\color{black} is commonly considered as it can easily be measured in far field \cite{Shim2018}.

%It can be computed as follows: 

\section{Cooling power as a function of quantum efficiency and bandgap}

We now proceed to calculate the optical radiative heat flux and electrical power densities of the TPX system. To do so, we need to define specific radiative properties and make certain assumptions about the system components. For our analysis, we first use a simplified model representing III-V materials to identify the optimal bandgap energy for cooling purposes. In the following section, the LED temperature is set to 290 K and the PV cell to 300 K.
The permittivity is taken to be $\varepsilon = 10 + i$ for both components of the system, and corresponds to a typical real part of the permittivity for III-V materials. For the quantum efficiency, we consider two values: \color{black}$\eqeled \:$\color{black} = 1 and 0.95. The second value was chosen since it may represent a realistic estimate for a GaAs LED. % When computing the \color{black}$\eqeled$\color{black}  using  heterostructures mainly made of AlGaAs, and the drift-diffusion CRESCENT-1D code \cite{Legendre2025Crescent1D} and  known expressions for Auger code, SRH and surface mechanisms, values close to 0.99 can be obtained.
When computing \color{black}$\eqeled \:$\color{black} from known expressions for Auger and SRH mechanism, values close to 0.99 can be obtained. Experimentally, values around 0.9 have been reported but those  depend very strongly on surface nonradiative recombinations. As a result, 0.95 provides a realistic estimate.
\color{black}
We can now evaluate the performance of different III-V-like materials in the TPX system, focusing particularly on identifying the effect of bandgap energies and \color{black}$\eqeled \:$\color{black} change for cooling applications.

\begin{figure}[t]
	\centering
	\captionsetup[subfigure]{position=top, labelfont=bf,textfont=normalfont,singlelinecheck=off,justification=raggedright,
		belowskip=-0pt,aboveskip=0pt}
	\begin{subfigure}{0.495\textwidth}
		\caption{}
		\includegraphics[width=\textwidth]{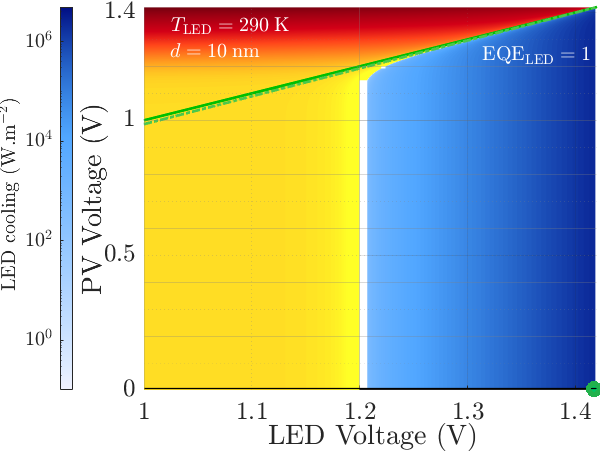}
		\label{fig:QcIII-VIQE1}
	\end{subfigure}
	\begin{subfigure}{0.495\textwidth}
		\caption{}
		\includegraphics[width=\textwidth]{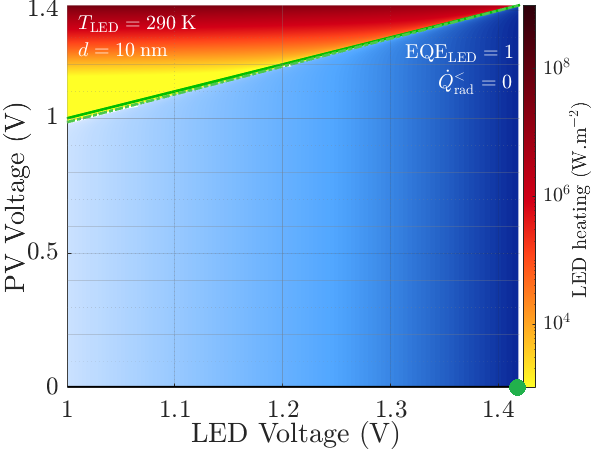}
		\label{fig:QcIII-VIQE1Nosub}
	\end{subfigure}
	\begin{subfigure}{0.495\textwidth}
		\caption{}
		\includegraphics[width=\textwidth]{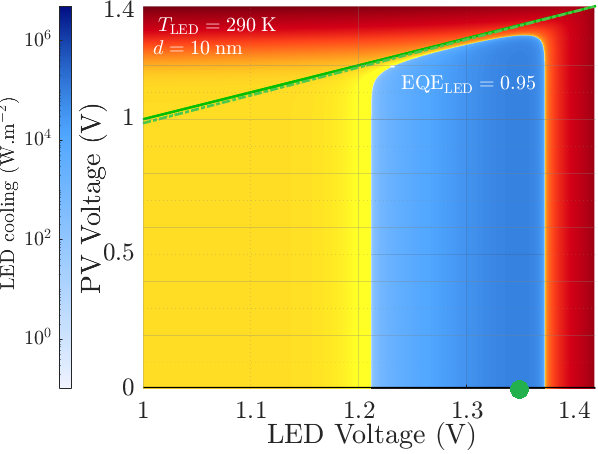}
		\label{fig:QcIII-VIQE95}
	\end{subfigure}
	\begin{subfigure}{0.495\textwidth}
		\caption{}
		\includegraphics[width=\textwidth]{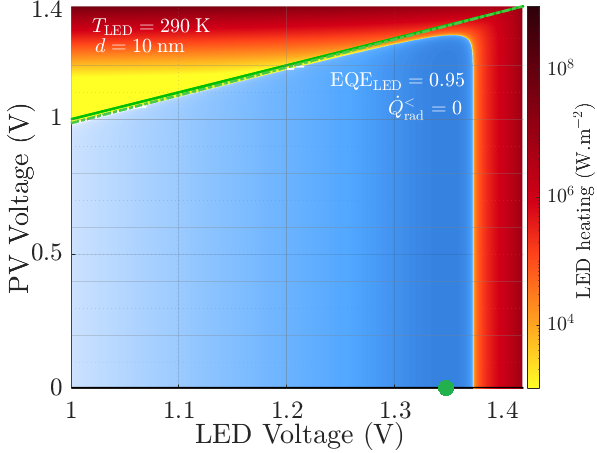}
		\label{fig:QcIII-VIQE95Nosub}
	\end{subfigure}
	
	\caption{ Cooling power as a function of LED and PV cell voltage biases for an LED and PV cell with dielectric constant $\varepsilon = 10 + i$. Cooling of LED occurs in blue area and heating occurs in the red area. Solid green line stands for $\Uled$ = $\Upv$ and dashed green line corresponds to the equality of the generalized Bose-Einstein distribution. Cooling power is given for: (a) \color{black}$\eqeled$\color{black} = 1 and including sub-bandgap radiative heat flux, (b) \color{black}$\eqeled$\color{black} = 1 and excluding sub-bandgap radiative heat flux, (c) \color{black}$\eqeled$\color{black} = 0.95 and including sub-bandgap radiative heat flux and (d) \color{black}$\eqeled$\color{black} = 0.95 and excluding sub-bandgap radiative heat flux. Green points show where maximum cooling power is obtained.}
	\label{fig:QcIII-V}	
\end{figure}

Figs. \ref{fig:QcIII-V}(a-d) show the cooling power as a function of LED and PV cell voltages for an arbitrarily-selected bandgap energy of 1.42 eV. From the top to bottom rows, the external quantum efficiency (\color{black}$\eqeled$\color{black}) decreases from 100\% to 95\%. The left column includes sub-bandgap radiative heat flux, while the right column sets this specific flux to zero ($\Qsub=0$), illustrating how efficient management of this quantity can benefit to the system. In these figures, one can distinguish two regions: the first one colored in blue shows the domain for which cooling occurs. The second one colored in yellow, orange or red highlights the biases for which LED heating happens. The solid line satisfies the condition $U_{\mathrm{LED}} = U_{\mathrm{PV}}$ and the dotted line below fulfils the condition at which the net flux is equal to zero \cite{Legendre2022}:
\begin{align}
	0 &= \int_{\omega_{gap}}^{+\infty} \left[\Theta(T\led,U\led,\omega)-\Theta(T\pv,U\pv,\omega)\right]\tau\s{tot}(\omega)  \:d\omega.
	\label{eq:NullCond}
\end{align}
Almost whole radiative flux is concentrated around the bandgap. From Eq. (\ref{eq:NullCond}) one gets
\begin{align}
	\Upv = \dfrac{T_{\mathrm{PV}}}{T_{\mathrm{LED}}}  \: U_{\mathrm{LED}} - \dfrac{E_{\mathrm{g}}}{e} \left( \dfrac{T_{\mathrm{PV}}}{T_{\mathrm{LED}}} - 1 \right) 
	\label{eq:BEeq1}
\end{align}
\begin{align}
 \qquad \qquad \quad = \left( \dfrac{1}{\text{COP}_\text{Carnot}} + 1\right)   \: U_{\mathrm{LED}} - \dfrac{E_{\mathrm{g}}}{e  \: \text{COP}_\text{Carnot}}.
\label{eq:BEeq}
\end{align}
The highest cooling power is achieved when the LED voltage matches the bandgap energy ($E\s{g} = e\Uled$) with a \color{black}$\eqeled$\color{black} of 1 (Fig. \ref{fig:QcIII-VIQE1}).  

Comparing Figs. \ref{fig:QcIII-VIQE1} and \ref{fig:QcIII-VIQE95} (same color scales), it is found that reducing \color{black}$\eqeled \:$\color{black} results in a lower maximum cooling power. This is due to an increase in the required external power to maintain the same radiative heat exchange, which partly converts to heat, up to a point where cooling is no longer achievable. 
As a result, the green markers, indicating the condition for maximum cooling power, shift towards lower LED voltages as \color{black}$\eqeled \:$\color{black} drops. As cooling power scales exponentially with LED bias ($U_{\mathrm{LED}}$), a lower bias can lead to a significant reduction in cooling power by two orders of magnitude.
Another feature of interest, highlighted by moving from the left (Figs. \ref{fig:QcIII-VIQE1} and \ref{fig:QcIII-VIQE95}) to the right column (Figs. \ref{fig:QcIII-VIQE1Nosub} and \ref{fig:QcIII-VIQE95Nosub}), is that suppressing sub-bandgap radiative flux allows cooling at lower voltage levels. The reason for neglecting sub-bandgap radiation is to provide an estimate of the maximum cooling power achievable. Such suppression could be implemented with a frequency-selective structure that would reflect all radiation below the bandgap while keeping practically untouched the above bandgap contribution.
Finally, note that biasing (powering) the PV cell increases the flux going from the PV cell to LED, thus decreasing the cooling power up to a point where refrigeration is no longer possible. The inclusion of a PV cell enables also photon conversion to electricity, therefore reducing the need for external electrical power and improving the COP. This effect cannot be seen on these figures where only cooling power is represented.

\begin{figure}[t]
	\centering
	\captionsetup[subfigure]{position=top, labelfont=bf,textfont=normalfont,singlelinecheck=off,justification=raggedright}
	\begin{subfigure}{0.49\textwidth}
		\caption{}
		\includegraphics[width=\textwidth]{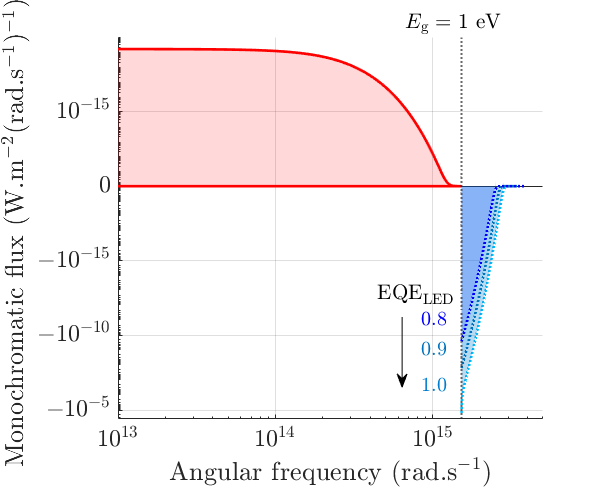}
		\label{fig:SpectraEgvar}
	\end{subfigure}
	\begin{subfigure}{0.49\textwidth}
		\caption{}
		\includegraphics[width=\textwidth]{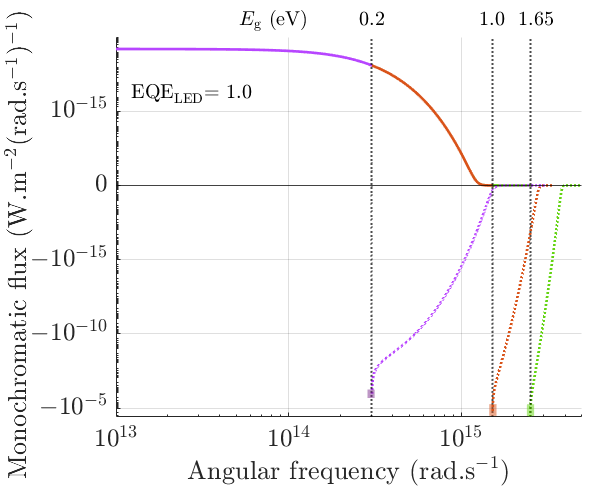}
		\label{fig:SpectraQEvar}
	\end{subfigure}
	\begin{subfigure}{0.49\textwidth}
	\caption{}
	\includegraphics[width=\textwidth]{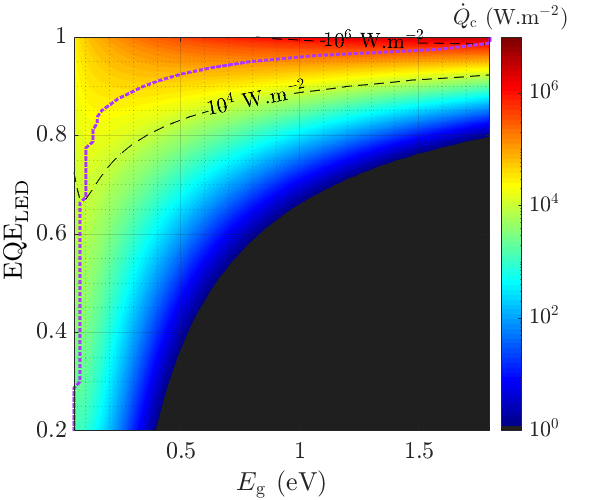}
	\label{fig:Cooling power EgQE}
\end{subfigure}
\begin{comment}
\begin{subfigure}{0.48\textwidth}
	\caption{}
	\includegraphics[width=\textwidth]{Fig2d.png}
	\label{fig:Cooling power EgQE cut}
\end{subfigure}
(d) Cooling power as a function of QE for each bandgap $\Eg$.
\end{comment}
	\caption{Maximum cooling power achievable as a function of quantum efficiency and bandgap energy for $d$ = 10 nm for a dielectric constant $\varepsilon = 10 + i$ in the LED and the PV cell\color{black}. (a) Monochromatic radiative flux for \color{black}$\eqeled$\color{black} = 1 and different values of bandgap energy. (b) Monochromatic radiative flux for $\Eg = 1$ eV and different values of \color{black}$\eqeled$\color{black}. Heating and cooling are displayed by red and blue areas, respectively. (c) Cooling power for optimized biases and $\Qsub$ = 0 as a function of bandgap energy and \color{black}$\eqeled$\color{black}. The purple line indicates the maximum cooling power for a given \color{black}$\eqeled$\color{black}. }
	\label{fig:Cooling power EgQE bloc}
\end{figure}

Let us now analyze spectral contribution to the radiation. Fig. \ref{fig:Cooling power EgQE bloc} investigates the effect of the critical parameters: quantum efficiency and bandgap energy. Fig. \ref{fig:SpectraEgvar} shows the net monochromatic radiative flux exchanged between a cold LED emitter and a room-temperature PV cell receiver for three different bandgap energies: 0.2 eV (purple), 1.0 eV (orange) and 1.65 eV (green). \color{black}$\eqeled \:$\color{black} is fixed at 1. The net cooling power is obtained by integrating the area below the curve where the contribution below the bandgap is set positive (heating of the LED since flux always goes from hot PV cell to cold LED) and the one above bandgap negative (cooling since the flux goes from the cold LED to the hot PV cell). Increasing the bandgap energy causes the electroluminescent peak to grow in absolute value leading to a higher cooling power since for higher \color{black}$\eqeled$\color{black}, higher bias voltages can be applied. Note that the bandgap energy and dielectric permittivity dependencies on temperature are here neglected, which could impact the intensity and position of the electroluminescent cooling peaks.
Fig. \ref{fig:SpectraQEvar} shows the net monochromatic radiative flux exchanged between a LED and a PV cell for 3 different \color{black}$\eqeled \,$\color{black} values. Only absolute values above 10$^{-19}$ W.m$^{-2}$. rad$^{-1}$.s. are represented in the figure. Sub-bandgap radiation corresponds to $\Qsub$ and the electroluminescent peak corresponds to $\Qabove$ from (see Eqs.(\ref{eq:Qsub}) and (\ref{eq:Qabove})). The three different shades of blue correspond to different \color{black}$\eqeled \,$\color{black} values: 0.8 (dark blue), 0.9 and 1.0. The distance is $d = 10$ nm and $E\s{g} = 1$ eV. The biases are selected so that the cooling power is maximized for each \color{black}$\eqeled$\color{black}. Decreasing \color{black}$\eqeled \:$\color{black} by 0.2 decreases the intensity of the electroluminescent peak at the gap frequency by five orders of magnitude. This peak reduction leads to a drop of the cooling power by two orders of magnitude.

We now compute the cooling power as a function of both \color{black}$\eqeled \:$\color{black} and $E\s{g}$. Each point of the figure is obtained by computing $\Qabove$ while considering $\Qsub = 0$ and optimizing the bias applied to LED and PV cell, i.e. according to the extraction of the green points highlighted in Fig. \ref{fig:QcIII-V}. %It assumes that parasitic thermal radiation can be eliminated by some spectral filtering: 
The data provide thus an upper bound to the cooling power. For the sake of simplicity all cooling values below 1 W.m$^{-2}$ are represented in dark grey. %Fig. \ref{fig:Cooling power Eg\color{black}$\eqeled$\color{black} cut} shows the evolution of $\Qc$ vs \color{black}$\eqeled$\color{black} for each fixed value of $E\s{g}$. It describes the evolution of the optimized cooling power with \color{black}$\eqeled$\color{black} by first fixing the bandgap energy. 
For all $E\s{g}$ the maximum is obtained for \color{black}$\eqeled$\color{black} = 1 as expected. For a bandgap energy of 0.30 eV, an increase of \color{black}$\eqeled \:$\color{black} from 80 \% to 100 \% results in a rise of a  four orders of magnitude of the cooling power. For a bandgap energy of 1.65 eV, an increase of \color{black}$\eqeled$\color{black} from 80 \% to 100 \% results in a rise of   beyond ten orders of magnitude of the cooling power. 
%In Fig. \ref{fig:Cooling power EgQE}, we chose to represent the bias-optimized cooling power according to the extraction of the green points highlighted in Fig. \ref{fig:QcIII-V} with $\Qsub = 0$, to show the expected maximum cooling power reachable for given QE, bandgap energy and with $\varepsilon = 10 + i$. Including sub-bandgap radiative flux restricts the situation in which cooling is possible. More precisely, cooling powers below $100$ \CoolDens$ \:$ are impossible to get as sub-bandgap radiative heat flux is too intense for low QEs. 
\color{black} One can distinguish therefore two optimal configurations. Considering a low \color{black}$\eqeled$\color{black} (\color{black}$\eqeled$\color{black} $\le$ 85 \%), it is advantageous to have a low bandgap-energy material. At high \color{black}$\eqeled$\color{black}, it is the opposite.  %Maximum cooling power can be reached when $E\s{g}$ = e$\Uled$ and QE = 1 with the temperature difference and gap size mentioned before.
Interestingly, there is an optimum bandgap energy for each quantum efficiency. 
 
Summarizing, the analysis demonstrates that optimizing the cooling power in a TPX system depends significantly on the interplay between bandgap energy and \color{black}$\eqeled$\color{black}. %For high QE values (close to 1), cooling power increases dramatically, especially at higher bandgaps, where it can reach six orders of magnitude more than for lower QE values. Conversely, at low QE, larger bandgaps reduce the cooling power as a greater proportion of electron-hole pairs do not recombine radiatively. Note that the density of photons going from the PV cell to the LED is constant in these test cases, and to achieve maximum cooling power no bias is applied to the PV cell. 
The need for large \color{black}$\eqeled \,$\color{black} and a high bandgap energy point towards III-V materials to maximize cooling performance in TPX systems, as they are known to possess high \color{black}$\eqeled \:$\color{black} with an already significantly high bandgap. 

\section{Cooling conditions and performances using a gallium arsenide based thermophonic device}

If the available \color{black}$\eqeled \:$\color{black} is below 80\%, low-bandgap materials have to be considered and the cooling power can reach $10^{4}$ \CoolDens.
This threshold is already reachable by commercialized Peltier module technology \cite{Zhao2014}. As a result, we chose to evaluate the performances of a gallium arsenide (GaAs) based realistic TPX system, as this alloy can reach high \color{black}$\eqeled$\color{black} \cite{Lastras-Martinez1978,Hwang1972,Vernon1992,Zayan2014,Madhusoodhanan2022}. It has a bandgap $E\s{g}$ at room temperature of about 1.42 eV \cite{Blakemore1982}. In this configuration, the expected cooling power could be of the order of magnitude of $10^6$ \CoolDens, which surpasses thermoelectrics. It has to be reminded the bandgap energies of emitter and receiver have to be matched in order to achieve optimum optical coupling \cite{Florescu2007,Bowman2021, Lopez2021,Wang2019}. As an increase of temperature leads to a decrease of the material bandgap \cite{Gonzalez-Cuevas2006,GonzalezCuevas2007}, we need to modify the composition of the PV cell to reduce the bandgap energy mismatch. It is better to avoid using a ternary alloy for the LED since electrical performances of such LEDs are known to be worse than for pure GaAs. The decrease of bandgap with temperature follows the semi-empirical Varshni law \cite{Varshni1967}: \vspace*{-.5cm}
\begin{equation}
	\centering
	E_{g}(T) = E_{g}(T = 0  \: \mathrm{K}) - \dfrac{\alpha T^{2}}{T + \beta}.
	\label{eq:Varshi}
\end{equation}
Specifically, for GaAs, $\Eg$ $(T = 0 $ K$) = 1.52$ eV, $ \alpha = 0.55$ meV.$\text{K}^{-1} $ and $\beta = 225$ K. To counteract the bandgap shift for higher temperature, a small proportion of aluminium is therefore incorporated in the PV cell. Our system is therefore made of an LED made of GaAs and a PV cell made of Al$_{\mathrm{x}}$Ga$_{\mathrm{1-x}}$As. We set first the LED temperature and determine its bandgap energy. We then compute the fraction of aluminium $x$ needed so that the PV cell bandgap energy matches that of the LED, using the model of interband transition provided in Ref. \cite{GonzalezCuevas2007}. Once the bandgap energy is known, we can determine the permittivity of both LED and PV cell. For the GaAs LED we use the data from \cite{GonzalezCuevas2007} for the permittivity above bandgap ($[\omega\s{gap}, +\infty[$) and \cite{Adachi1994} for the permittivity below bandgap ($ [ 0,\omega\s{gap}] $). For the permittivity of the Al$_{\mathrm{x}}$Ga$_{\mathrm{1-x}}$As PV cell above bandgap we take the data from \cite{GonzalezCuevas2007} and below bandgap from \cite{Lukes1989}. 
% It is now possible to compute more realistic cooling power maps using different LED biases $\Uled$ and PV biases $\Upv$. This way, we can determine the required biases needed to achieve optimum cooling with respect to different gap distances, LED temperatures and \color{black}$\eqeled$\color{black}s. 
Using this method, we find that the required amount aluminum fraction to put into a 300 K PV cell for a 290 K GaAs LED is $0.37 \%$ and for a 250 K LED is $1.81\%$. Note that in the case of GaAs, there is a clear spectral separation between the sub-bandgap radiation $\Qsub$, which is due to thermal radiation, and the above-bandgap one $\Qabove$, which is due to electroluminescence only.\color{black}
\subsection{Optimal cooling power and impact of thermal radiation}
\begin{figure}[!ht]
	\captionsetup[subfigure]{position=top, labelfont=bf,textfont=normalfont,singlelinecheck=off,justification=raggedright,
		belowskip=-30pt}	
	\begin{subfigure}{0.475\textwidth}
		\caption{} 		
		\includegraphics[width=\textwidth]{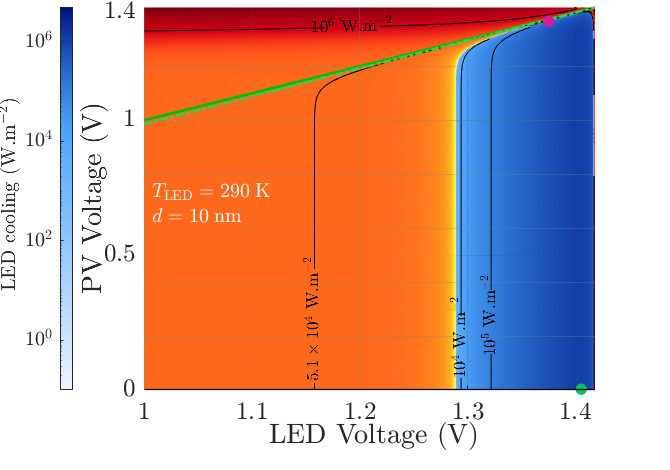}
		\label{fig:Cooling map 290 K 10 nm}
	\end{subfigure} 
	\begin{subfigure}{0.475\textwidth}
		\caption{} 		
		\includegraphics[width=\textwidth]{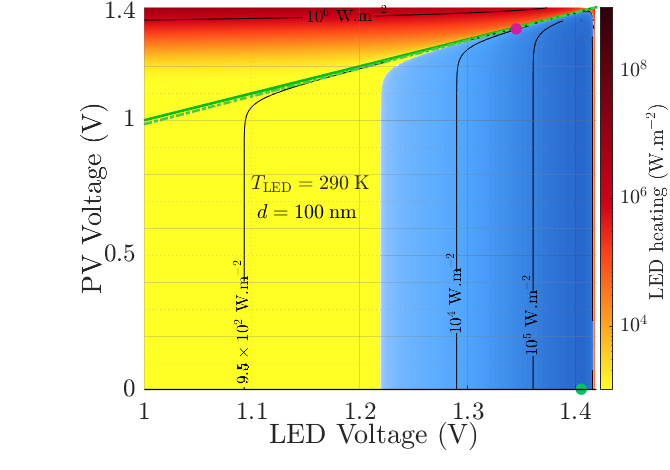}
		\label{fig:Cooling map 290 K 100 nm}
	\end{subfigure} 
	\begin{subfigure}{0.475\textwidth}
		\caption{} 		
		\includegraphics[width=\textwidth]{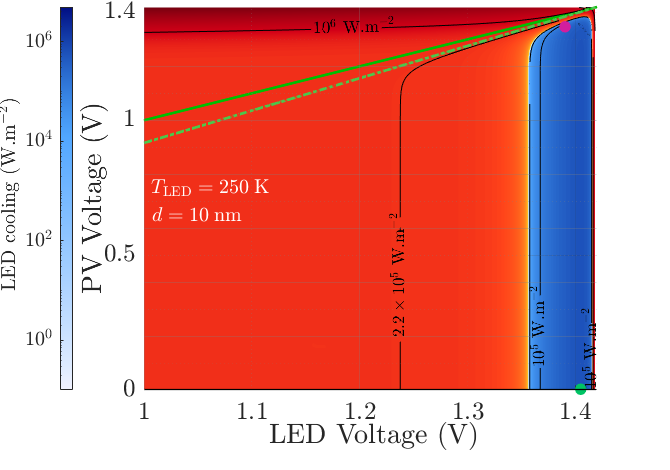}
		\label{fig:Cooling map 250 K 10 nm}
	\end{subfigure} 
	\begin{subfigure}{0.475\textwidth}
		\caption{} 		
		\includegraphics[width=\textwidth]{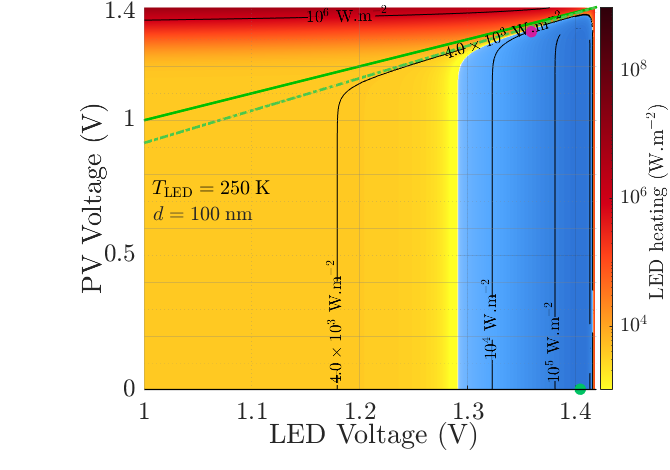}
		\label{fig:Cooling map 250 K 100 nm}
	\end{subfigure}
	\begin{subfigure}{0.45\textwidth}
		\caption{} 		
		\includegraphics[width=\textwidth]{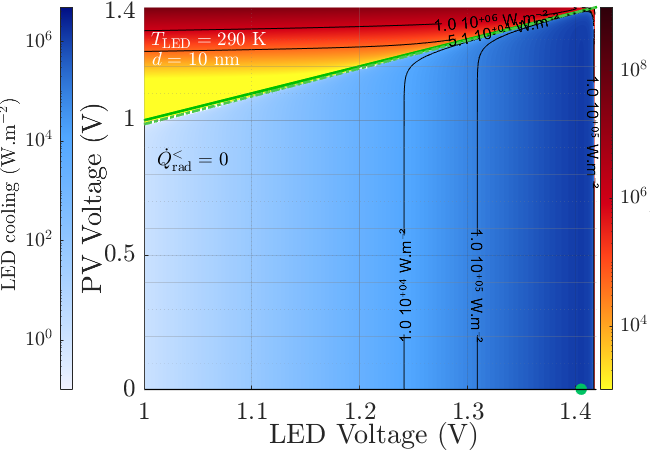}
		\label{fig:Cooling map 290 K 10 nm Nosub}
	\end{subfigure} \vspace{-15pt}
	\caption{Cooling power as a function of LED and PV cell voltage biases for a GaAs LED and a matched-bandgap AlGaAs PV cell at the radiative limit\color{black}. Cooling of LED occurs in blue area and its heating occurs in the red area. Solid green line stands for $\Uled$ = $\Upv$ and dashed green line corresponds to the equality of the generalized Bose-Einstein distribution. Cooling power is given for: (a) $\Tled=290$ K  and $d = 10$ nm, (b) $\Tled=290$ K  and $d = 100$ nm, (c) $\Tled=250$ K  and $d = 10$ nm, (d) $\Tled=250$ K  and $d = 100$ nm, (e) $\Tled=290$ K , $d = 10$ nm and excluding sub-bandgap radiative heat flux. Green (resp. pink) points show where maximum cooling power (resp. COP) are obtained.  }
	\label{fig:Cooling map T}
\end{figure}

Fig. \ref{fig:Cooling map T} shows the cooling power map for two LED temperatures (250 and 290 K) and two gap distances (10 and 100 nm). To appreciate the optimized cooling possibilities offered by this system, $\Qsub$ has been set to 0 in Fig. \ref{fig:Cooling map 290 K 10 nm Nosub}. From Fig. \ref{fig:Cooling map 290 K 10 nm} to Fig. \ref{fig:Cooling map 250 K 100 nm}, we include thermal radiation (i.e. $\Qsub \ne 0$), leading to a reduced cooling domain and thus imposing stricter cooling conditions. Above $\Uled = \frac{\Eg}{e}$, there is no cooling power.
Reducing the LED temperature from 290 K to 250 K (as seen between Fig. \ref{fig:Cooling map 290 K 10 nm} and Fig. \ref{fig:Cooling map 250 K 10 nm}) increases thermal radiation, shown by a darker red shade. This rise in thermal radiation results from an increase in the modified Bose-Einstein distribution difference, $\Theta(T_{\mathrm{PV}},U_{\mathrm{PV}},\omega) - \Theta(T_{\mathrm{LED}},U_{\mathrm{LED}},\omega)$. Since thermal radiation is independent of the biases applied to the LED and PV, it serves as an initial barrier to achieving cooling power at moderate LED biases (around 1.15 V). However, the increased distribution difference also reduces electroluminescent radiation, making it more challenging to overcome thermal radiation at lower LED temperatures. Consequently, a higher LED bias is required to achieve cooling. As the temperature difference between LED and PV cells grows, the boundary of the cooling power zone (shown by the green dashed line) shifts, limiting the range of conditions where cooling power can be achieved.\\
Between Fig. \ref{fig:Cooling map 290 K 10 nm} and Fig. \ref{fig:Cooling map 290 K 100 nm} and between Fig. \ref{fig:Cooling map 250 K 10 nm} and Fig. \ref{fig:Cooling map 250 K 100 nm}, the gap distance has been increased from 10 nm to 100 nm. Near-field effects are reduced leading to a smaller $\Qsub$ and also a lower $\Qabove$, i.e. to a reduction of both thermal and electroluminescent radiation. Nevertheless, between the two, thermal radiation decreases faster, as thermal radiation is strongly related to surface modes with strong distance dependence at very short distances, and electroluminescent radiation is linked to frustrated modes, which level off at these short distances. As a result, cooling power can be achieved at more flexible conditions as $\Qsub$ decreases from 5.1 $\times$ $10^{4}$ W.m$^{-2}$ in Fig. \ref{fig:Cooling map 290 K 10 nm} to 9.5 $\times$ $10^{2}$ W.m$^{-2}$ in Fig. \ref{fig:Cooling map 290 K 100 nm}. For $U\pv$ = 0, cooling can be achieved for $U\led \gtrsim 1.29$ V for $d = 10$ nm whereas it is achieved for $\Uled \gtrsim 1.22$ V for $d = 100$ nm.  Also, by comparing Fig. \ref{fig:QcIII-VIQE1} and Fig. \ref{fig:Cooling map 290 K 10 nm}, one can see that the sub-bandgap radiative heat flux can be larger for a more realistic material. This is due to the fact that the phonon-polariton contribution is strong in GaAs-based materials\color{black}. All these results are obtained for ideal conversion of electron-hole pairs into photons (\color{black}$\eqeled$\color{black}$=1$). 
\newpage
\subsection{Cooling performances vs distance}
\begin{figure}[t]
	\centering
	\captionsetup[subfigure]{position=top, labelfont=bf,textfont=normalfont,singlelinecheck=off,justification=raggedright,
	belowskip=-30pt}	
	\begin{subfigure}{0.495\textwidth}
		\caption{}
		\includegraphics[width=\textwidth]{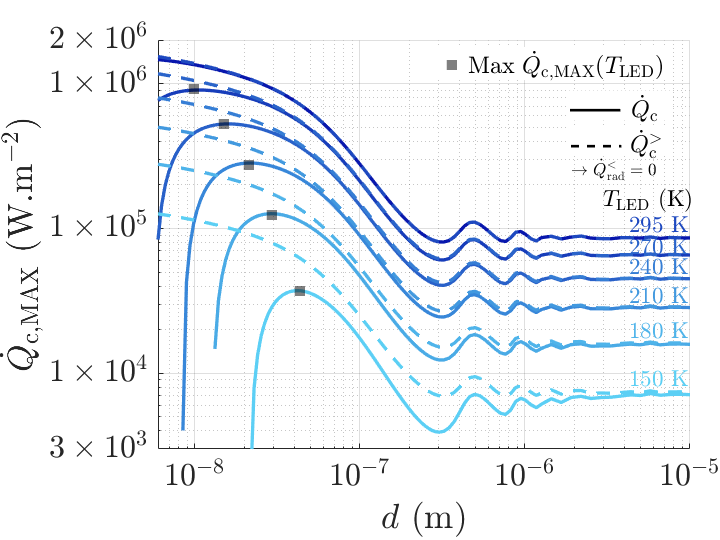}
		\label{fig:Optimised cooling powerdT}
	\end{subfigure}
	\begin{subfigure}{0.495\textwidth}
		\caption{}
		\includegraphics[width=\textwidth]{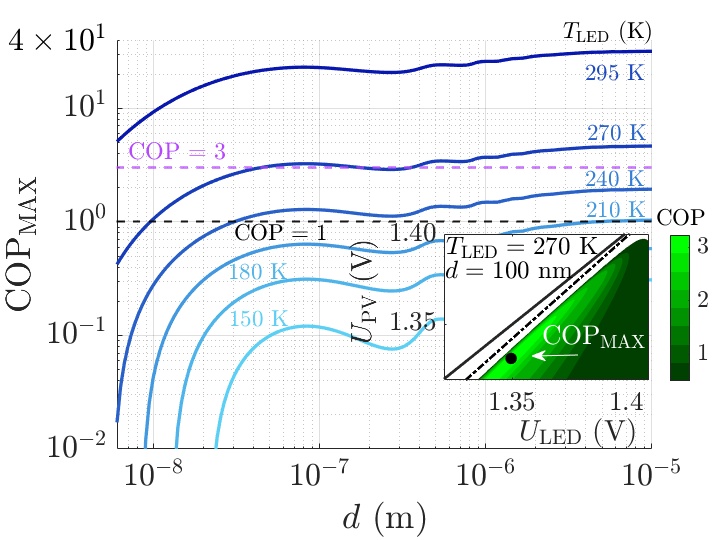}
		\label{fig:Optimised SCOPdT}
	\end{subfigure}
	\begin{subfigure}{0.495\textwidth}
		\caption{}
		\includegraphics[width=\textwidth]{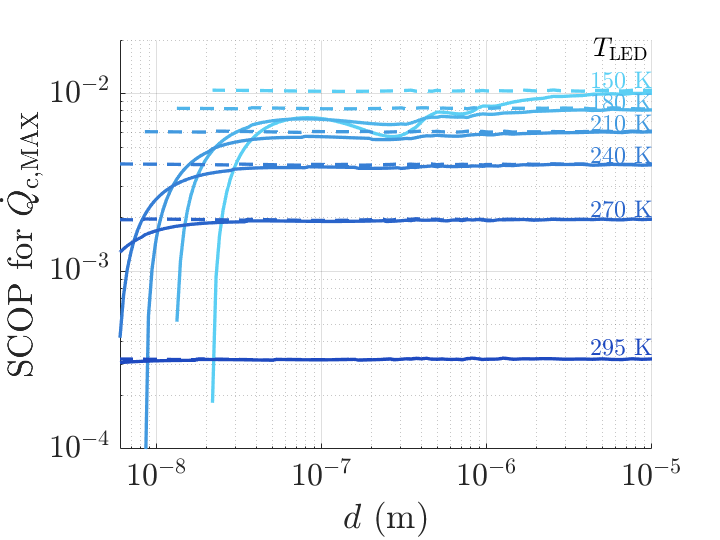}
		\label{fig:COP for Optimised Qc}
	\end{subfigure}
	\begin{subfigure}{0.495\textwidth}
		\caption{}
		\includegraphics[width=\textwidth]{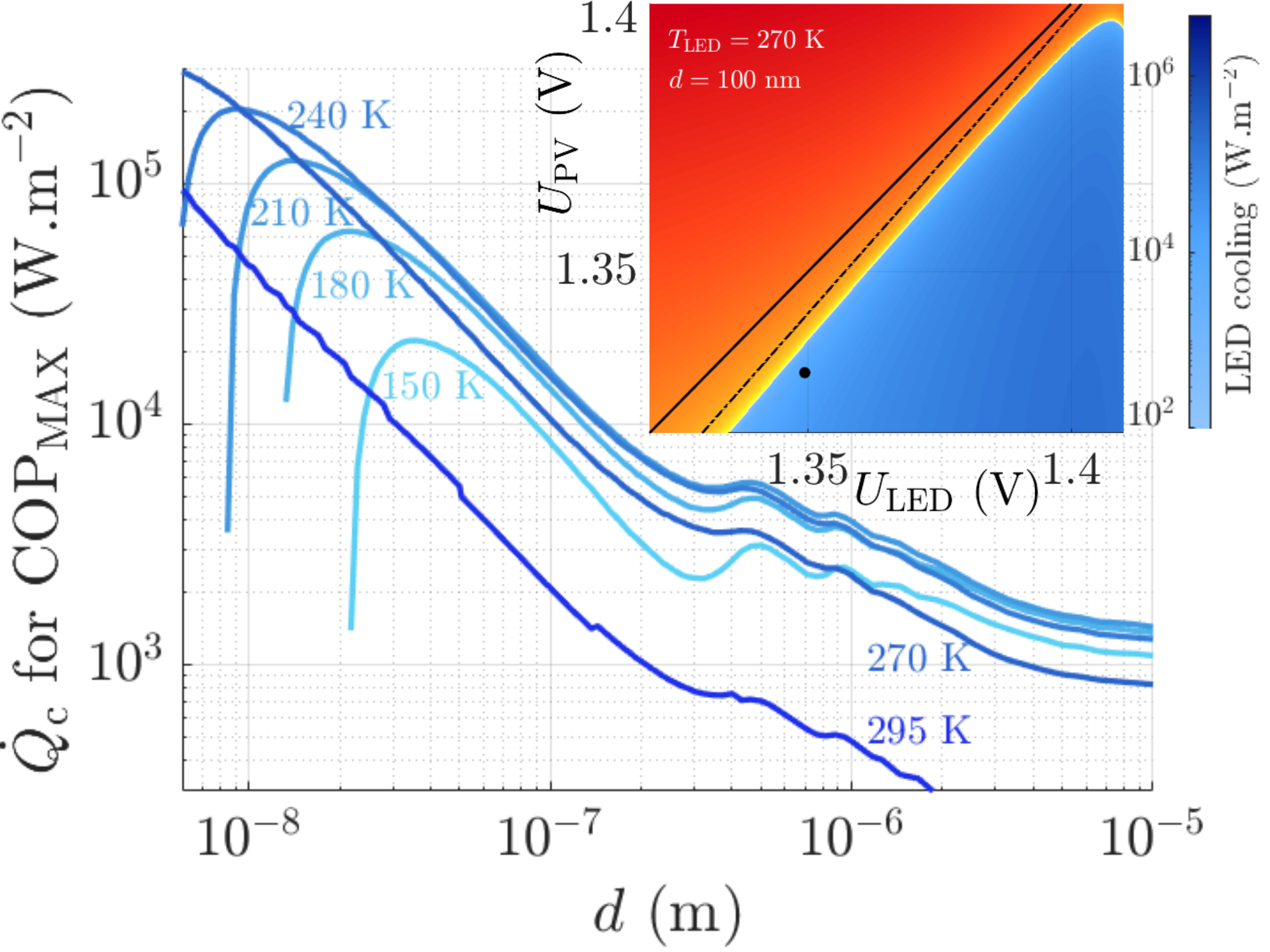}
		\label{fig:Qc for Optimised COP}
	\end{subfigure}
	\caption{Performances for optimized cooling power (a,c) and optimized coefficient of performance (b,d) as a function of LED temperature $\Tled$ and gap size $d$ for a GaAs LED and a AlGaAs PV cell at the radiative limit\color{black}. The dashed lines correspond to the case where no sub-bandgap radiation is exchanged ($\Qsub = 0$) while the solid lines include $\Qsub$. Each color stands for a different LED temperature. In (b), COP is displayed in the insert as a function of LED and PV biases for an LED temperature of $270$ K and a gap distance of $100$ nm, with maximum COP being highlighted by a black dot. In (d), cooling power is displayed in the insert as a function of LED and PV biases for the same parameters, focusing on the region where COP $\ge$ 0.5. The  cooling power at maximum COP is represented by a black dot. In (b), the horizontal dashed purple line indicates a typical value for refrigerators.  \cite{Silva-Romero2024}\color{black}}
	\label{fig:OptimisedCore}
\end{figure}
We now look in more details at the effect of near field on performances. To do so, Fig. \ref{fig:OptimisedCore} displays as a function of distance both the optimized cooling power and the scaled coefficient of performance. % (SCOP), which is the ratio of the actual COP to its theoretical upper bound. 
As can be expected, cooling power is larger for small LED-PV temperature difference as shown in Fig. \ref{fig:Optimised cooling powerdT}. Let us consider first dashed lines for which $\Qsub = 0$.
We can distinguish three regimes on each curves: for $d > 2 $ \textmu m, we are in far field for electroluminescence. As only propagating modes are exchanged in the system, the optimized cooling power is constant. For $d \ll \lambda_g=\frac{hc}{E_\mathrm{g}}$, we are in near field and electroluminescent radiation is enhanced due to evanescent modes contribution, resulting in a ten-fold increase of the cooling power. The increase is due to frustrated modes (the typical divergence due to surface modes is not observed in this distance range). For $ 300  \: \mathrm{nm} < d < 2  \: $\textmu m , some optical interferences from propagating modes lead to oscillations in the cooling power. 
For each dashed line representing the case where $\Qsub = 0$ corresponds a solid line in which $\Qsub$ is accounted for, which underlines the issue of thermal radiation in the TPX refrigerators. For $T_{\mathrm{LED}} > 210$ K and  $d > 200 $ nm, thermal radiation has a negligeable impact in our TPX system, as the two lines are superimposed. For short distances and $T_\mathrm{{LED}} < 270$ K, the optimized cooling power drops, due to thermal radiation increasing at a higher pace than electroluminescent radiation. Therefore, there is an optimal distance in which the maximum cooling power for a given LED temperature is reached. 
As expected, optimizing the cooling power in this system leads to a poor energy efficiency if thermal radiation is accounted for as shown in Fig. \ref{fig:COP for Optimised Qc}. In this scenario, the PV cell is not producing any power and a strong power needs to be supplied to the LED, thus resulting in a need for a strong external electrical power $\Pelec$, and finally a low SCOP. In fact, all displayed SCOP values including thermal radiation are smaller than 1 \%. Again, it is seen that a strong thermal-radiation management could limit the degradation of the SCOP. Considering maximum COP as a function of $\Tled$ and $d$ in Fig. \ref{fig:Optimised SCOPdT}, the previous three regions can be observed. At $d \lesssim$ 60 nm, an increase of the SCOP can be seen when the gap size increases as the required amount of electrical power decreases faster than the cooling power. Notably, the local maximum COP at $d \approx 60$ nm is recovered for all LED temperatures when the LED–PV cell separation reaches the transition between the near-field and far-field regimes for electroluminescence ($d \ge 1$ \textmu m). The inset highlights the region where the maximum COP is attained for each computation of cooling power as a function of LED temperature and separation distance. The cooling power resulting for the energy efficiency optimization is displayed in Fig. \ref{fig:Qc for Optimised COP}. The darkest line represents $T_\mathrm{{LED}} = 295$ K and displays the overall lowest cooling power values. As the electroluminescent radiation from the PV cell to the LED is larger in this case, the cooling power is lower than in the optimised cooling-powercase. Having a high bias in the PV cell enables the conversion of electroluminescent power into electric power, thus reducing the amount of needed external electrical power. The combination of those two statements results in having a system with substantially higher energy efficiency, but with a lower cooling power \cite{Legendre2024}.
\begin{figure}[t]
	\centering
	\captionsetup[subfigure]{position=top, labelfont=bf,textfont=normalfont,singlelinecheck=off,justification=raggedright,
		belowskip=-0pt,aboveskip=0pt}
	\begin{subfigure}{0.45\textwidth}
		\caption{}
		\includegraphics[width=\textwidth]{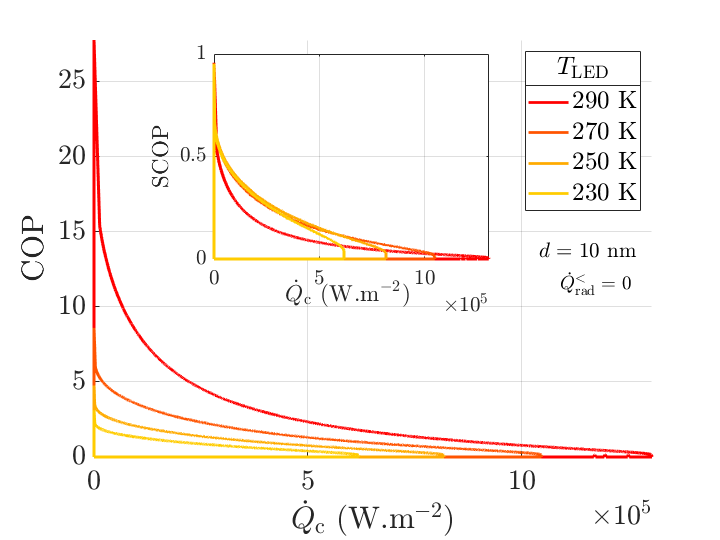}
		\label{fig:QcMAXGaAsIQE95}
	\end{subfigure}
	\begin{subfigure}{0.45\textwidth}
		\caption{}
		\includegraphics[width=\textwidth]{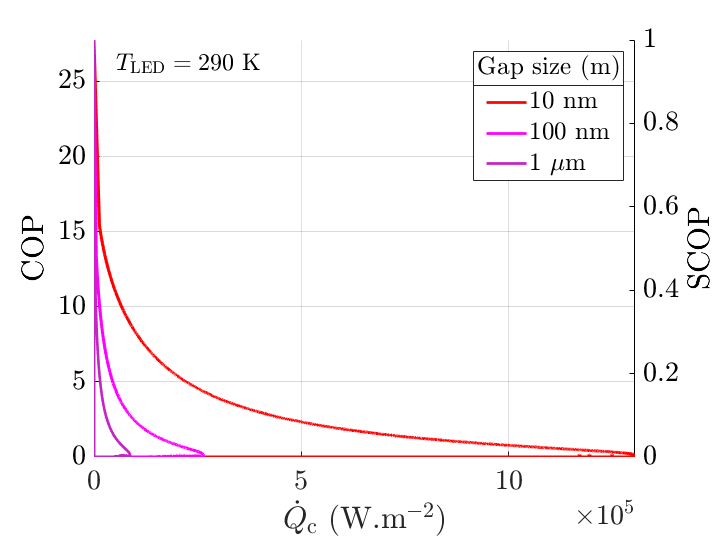}
		\label{fig:COPMAXGaAsIQE95}
	\end{subfigure}
		\begin{subfigure}{0.45\textwidth}
		\caption{}
		\includegraphics[width=\textwidth]{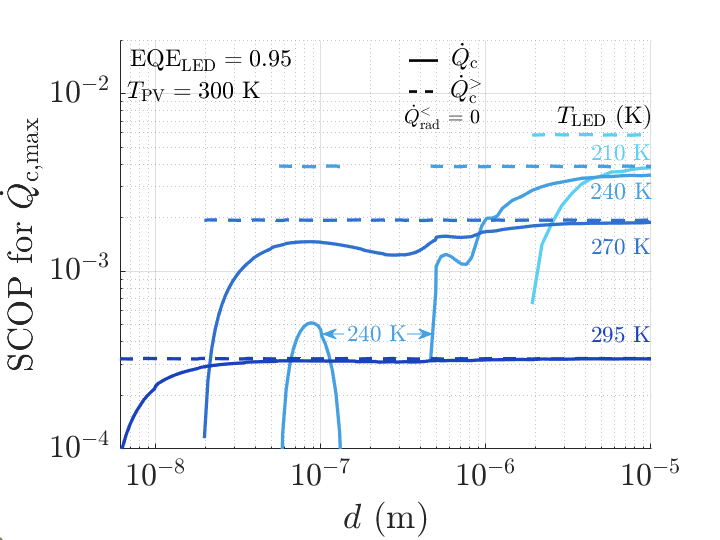}
		\label{fig:COP for Optimised Qc95}
	\end{subfigure}
		\begin{subfigure}{0.45\textwidth}
		\caption{}
		\includegraphics[width=\textwidth]{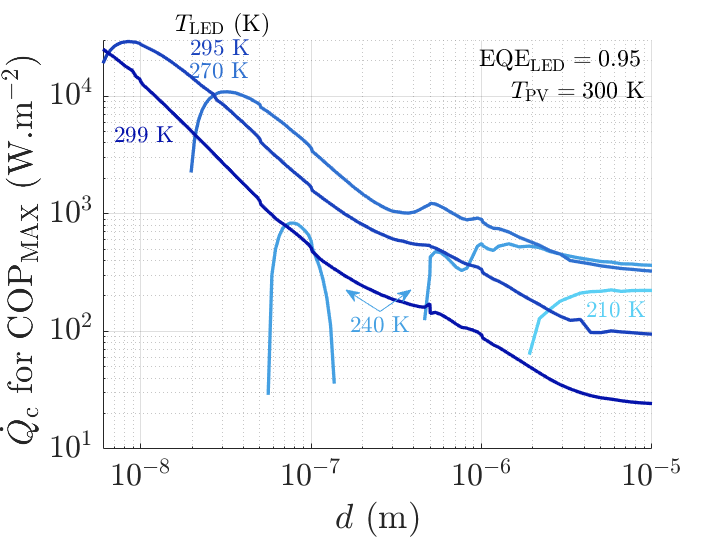}
		\label{fig:Qc for Optimised COP95}
	\end{subfigure}
	\caption{(a) Performances for optimized cooling power, and (c) its corresponding SCOP. (b) Optimized coefficient of performance, and (d) its  corresponding cooling power as a function of LED temperature $\Tled$ and gap size $d$ for a GaAs LED and a AlGaAs PV cell for \color{black}$\eqeled$\color{black} = 0.95. All solid lines include sub-bandgap radiation. The dashed lines in (a) and (c) correspond to the case where no sub-bandgap radiation is not exchanged ($\Qsub = 0$).\color{black} }
	\label{fig:GaAsIQE95}	
\end{figure}

 We now address the case of a reduced quantum efficiency, of 95\% (Fig. \ref{fig:GaAsIQE95}). Comparing Fig. \ref{fig:Optimised cooling powerdT} and \ref{fig:QcMAXGaAsIQE95}, we observe that reducing \color{black}$\eqeled \:$\color{black} leads to %at least a two-order-of-magnitude 
a strong drop in cooling power across all LED temperatures. In addition, it can be seen that cooling cannot be achieved for certain distance ranges. %The maximum temperature at which net cooling remains feasible is approximately 90 K.
%We can see that the maximum temperature in which cooling is achievable is at maximum 90 K. This is due to the fact that even though the QE is lower, the radiative heat flux attached to the system does not depend on QE whereas current density calculation does. More precisely, the sub-bandgap radiative heat flux remains the same, which impedes the system from cooling as greatly as with QE = 1.
% This performance degradation arises because, while QE affects current density, the associated radiative heat flux, and particularly sub-bandgap radiation, remains largely unchanged. As a result, the reduced QE leads to higher non-radiative losses, which limit cooling effectiveness compared to the QE = 1 case.
Fig. \ref{fig:COPMAXGaAsIQE95} shows the optimized coefficient of performance (COP) as a function of distance. The dashed line indicates a COP of 3 which is typical for conventional compression-based cooling systems \cite{Silva-Romero2024}\color{black}. As can be seen, the thermophotonic system does not reach this value under any conditions. Even reaching a COP of 1 proves difficult due to both lower \color{black}$\eqeled \,$\color{black} and high sub-bandgap radiative heat flux.  Fig. \ref{fig:COP for Optimised Qc95} is the counterpart of Fig. \ref{fig:COP for Optimised Qc} with \color{black}$\eqeled$\color{black} = 95\%, and shows comparable SCOP for temperature differences below 60 K and SCOP values dropping up to 50\% for higher temperature differences. Finally, the counterpart of Fig. \ref{fig:Qc for Optimised COP},  providing the cooling power at maximum COP, is Fig. \ref{fig:Qc for Optimised COP95}. For every bias-optimized COP points, the associated cooling power is lower by a ten-fold. Given the previous figures with a closer-to-real \color{black}$\eqeled$\color{black}, the GaAs-based cooling TPX system could operate with cooling power ranging up to $10^{4}$ \CoolDens $ \:$ and its COP can reach up to almost 0.3 for temperature differences above 30 K.  To sum up, such device without low sub bandgap transfer  management has limited viability\color{black}. 
%Fig SCOP
\subsection{Trade-off between cooling power and COP}
\begin{figure}[t]
	\captionsetup[subfigure]{position=top, labelfont=bf,textfont=normalfont,singlelinecheck=off,justification=raggedright}
	\begin{subfigure}{0.45\textwidth}
		\caption{}
		\includegraphics[width=\textwidth]{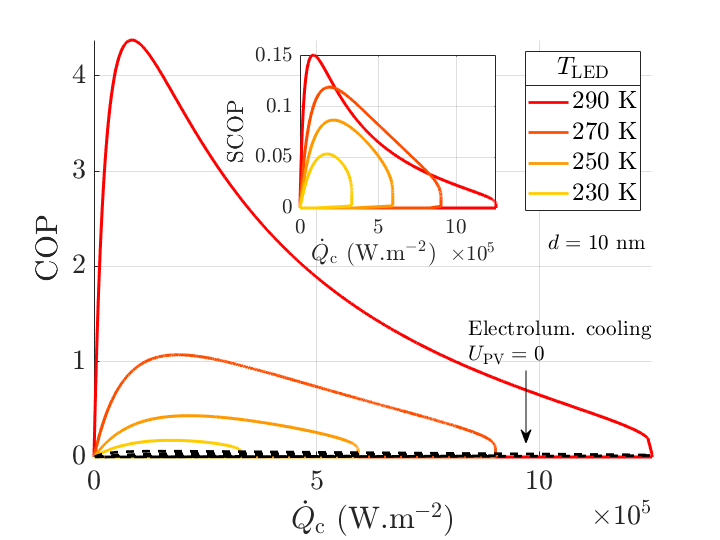}
		\label{fig:COP10nm RadLim}
	\end{subfigure}
	\begin{subfigure}{0.45\textwidth}
		\caption{}
		\includegraphics[width=\textwidth]{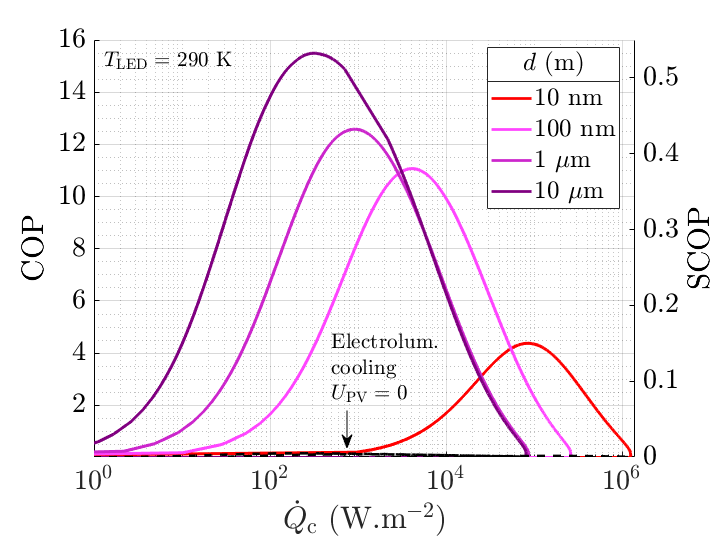}
		\label{fig:COP290K RadLim}
	\end{subfigure}
	\begin{subfigure}{0.45\textwidth}
		\caption{}
		\includegraphics[width=\textwidth]{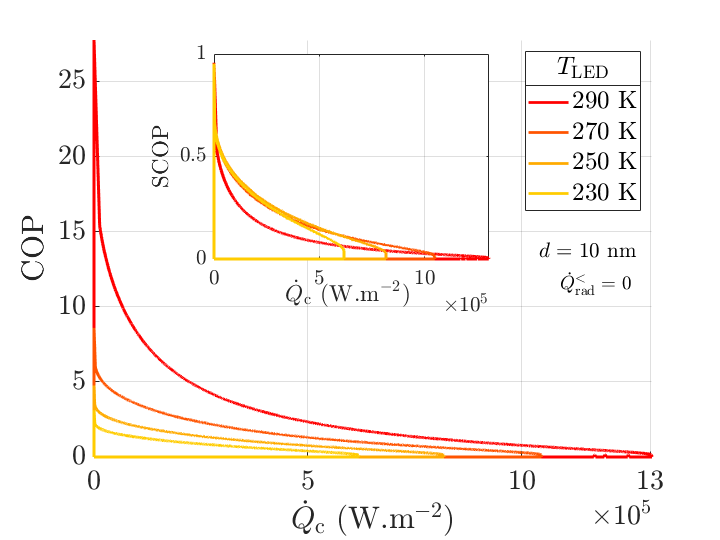}
		\label{fig:COP10nm NoSub}
	\end{subfigure}
	\begin{subfigure}{0.45\textwidth}
		\caption{}
		\includegraphics[width=\textwidth]{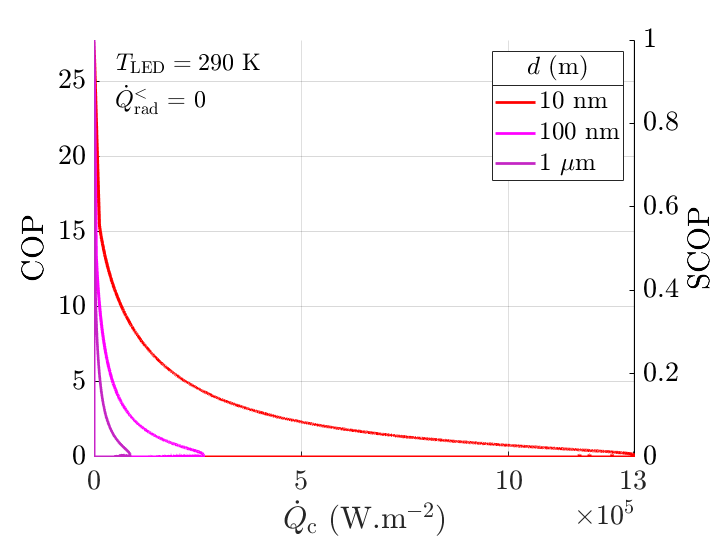}
		\label{fig:COP290K NoSub}
	\end{subfigure}
	\caption{ Performance plot of the near-field thermophotonic cooling system at the radiative limit (\color{black}$\eqeled$\color{black} = 1). Dashed line shows the performances of the TPX cooling device in the electroluminescent cooling configuration, i.e. with $\Upv = 0$ V. (a) Dependence on LED temperature including thermal radiation. (b) Dependence on distance including thermal radiation. (c) Dependence on LED temperature excluding thermal radiation. (d) Dependence on distance excluding thermal radiation.}
	\label{fig:COPSCOP QE1}
\end{figure}

 %Having established the methodology for computing cooling power in thermophotonic systems and characterized both its upper bounds and corresponding COPs, we now seek to synthesize these findings into a unified performance map. More specifically, we construct a parametric space plotting COP as a function of cooling power, capturing all achievable operating conditions.
We now provide performance plots, which give the relation between COP and cooling power. To do this, we extract from our prior computations in the $(U\led, U\pv)$ maps each operating point in which net cooling is possible, and compile all the associated cooling powers and their corresponding COPs. We then represent the envelope, i.e. the outer boundary of the possible phase space. %Our emphasis on cooling power originates from the scarcity of experimental points in the literature. Prioritizing the maximization of cooling power, followed by optimization of the photovoltaic bias to enhance efficiency, offers a framework to get the best from this relatively new dual optoelectronic engine.
Depending on the application, optimizing either the COP or the cooling power may be the priority. 
Metrics such as the product of COP and cooling power have been previously proposed \cite{Liao2019} in order to find a single interesting operating point. Most of the time, our analysis favors examining the maximum cooling as it enables comparison with other cooling systems. %and gives targets for interested experimenters and then look for optimization of COP near optimized LED bias. 
However, intermediate choices may be useful. The performance plots can be seen in Fig. \ref{fig:COPSCOP QE1}, which provides the COP and/or the SCOP including thermal radiation (below-bandgap radiation) as a function of cooling power. %By analyzing the envelope of points in terms of cooling power versus COP, we obtain a characteristic curve for each LED temperature and gap distance, providing an overview of the TPX system’s cooling performance.
%To produce these curves, we collect all points within the cooling region from previous cooling maps, then calculate the COP and derive the SCOP.
 In general, there is a trade-off: achieving high cooling power requires reducing the PV cell’s bias $\Upv$ in order to reduce the detrimental heat flow from the PV cell to the LED . Conversely, increasing the PV cell’s bias allows efficient conversion of electroluminescent radiation from the LED, especially at high \color{black}$\eqeled$\color{black}, and leads to a reduction of the needed external electrical power $\Pelec$, at the cost of a decrease in cooling power. % Therefore, even if cooling power is lower, the faster reduction in $\Pelec$ results in a higher COP. To sum up, Fig. \ref{fig:COPSCOP QE1} illustrates the trade-off between the cooling ability of the near-field TPX system and its COP. 
In Fig. \ref{fig:COP10nm RadLim}, since $\Qsub \ne 0$, a minimum electroluminescent radiation threshold must be exceeded for cooling to occur. This explains why low cooling power ($\Qc < 1$ \CoolDens) does not coincide with high COP ($\text{COP} > 3$). Thermal radiation can be considered as a heat leak of the hot reservoir towards the cold reservoir \cite{Pathria1993,Huleihil2011}. %Concentrating on Fig. \ref{fig:COP10nm RadLim}, the LED and PV cell biases can be adjusted to get a 
The best cooling power is $1.25 \times 10^{6}$ \CoolDens $ \:$ at a $10$ nm distance at $\Tled = 290$ K. Another option is to get the maximum COP of $\approx$ 4.38 with an associated cooling power of $8.42 \times 10^{4}$ \CoolDens $ \:$. %for the same distance and temperature difference. 
The main reason for the COP decrease when the LED temperature decreases is the increase of the  thermal radiation contribution, as higher temperature difference (i.e by lowering $\Tled$ in our computations) imposes greater Bose-Einstein factor difference. %In other words, a larger temperature difference induces an increase of $\Qsub$, which prevents cooling for lower $\Uled$ and $\Upv$ biases. In fact, as temperature difference decreases, the Bose-Einstein equality tends to behave as $\Uled = \Upv$, enabling the system to approach $\Pelec$ $\approx$ 0. 
As the distance between the two components increases, the maximum cooling power decreases %(compare Fig. \ref{fig:Cooling map 290 K 10 nm} and Fig. \ref{fig:Cooling map 290 K 100 nm}% or Fig. \ref{fig:Cooling map 250 K 10 nm} and Fig. \ref{fig:Cooling map 250 K 100 nm}
but COP and thus SCOP increase as seen in Fig. \ref{fig:COP290K RadLim}. %As less radiation is exchanged in the system, less thermal radiation is received by the LED, thus increasing the COP of the system. 
To compare with electroluminescent cooling, we include the results we obtain using $\Upv = 0$, which makes the TPX cooling system acting as a pure electroluminescent-cooling device. The cooling performances of the TPX device are not reduced. Oppositely, the obtained COP is close to zero, underlining that one main effect of the NF-TPX device is not to have better cooling performances but to enhance the sustainability of radiative solid-state cooling.

In Figs. \ref{fig:COPSCOP QE1}(c,d) we set $\Qsub = 0$, resulting in an improvement of the cooling power and a substantial step-up of the COP in comparison to the data of Figs. \ref{fig:COPSCOP QE1}a and \ref{fig:COPSCOP QE1}b. As can be seen in Fig \ref{fig:COP10nm NoSub}, the SCOP almost reaches 1 in all conditions in the inset of Fig. \ref{fig:COP10nm NoSub}. In those situations, as emission due to electroluminescence is unfortunately not monochromatic, some thermalisation losses are present in the system. This results in a SCOP approaching but distinct from 1. A SCOP of 1 could be reached for monochromatic radiation or with a very large bandgap \cite{Legendre2024}. In Fig. \ref{fig:COP290K NoSub}, it is seen that filtering thermal radiation allows reaching significant values of COP and large cooling power for a broad range of distances.
\color{black}

\begin{figure}[t]
	\captionsetup[subfigure]{position=top, labelfont=bf,textfont=normalfont,singlelinecheck=off,justification=raggedright}	
	\begin{subfigure}{0.45\textwidth}
		\caption{}
		\includegraphics[width=\textwidth]{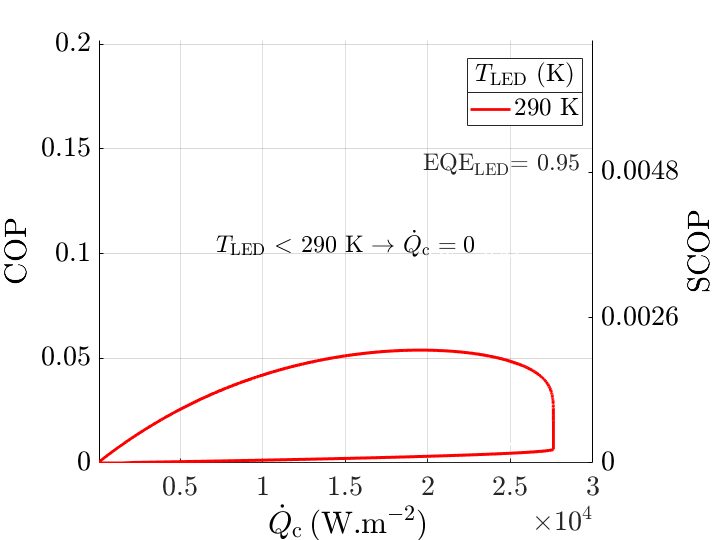}
		\label{fig:COP10nm QE095}
	\end{subfigure}
	\begin{subfigure}{0.45\textwidth}
		\caption{}
		\includegraphics[width=\textwidth]{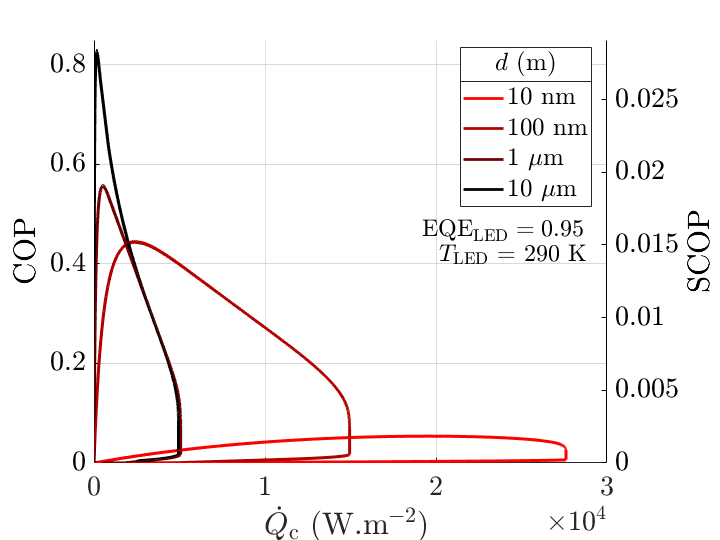}
		\label{fig:COP290K QE095}
	\end{subfigure}
	\begin{subfigure}{0.45\textwidth}
		\caption{}
		\includegraphics[width=\textwidth]{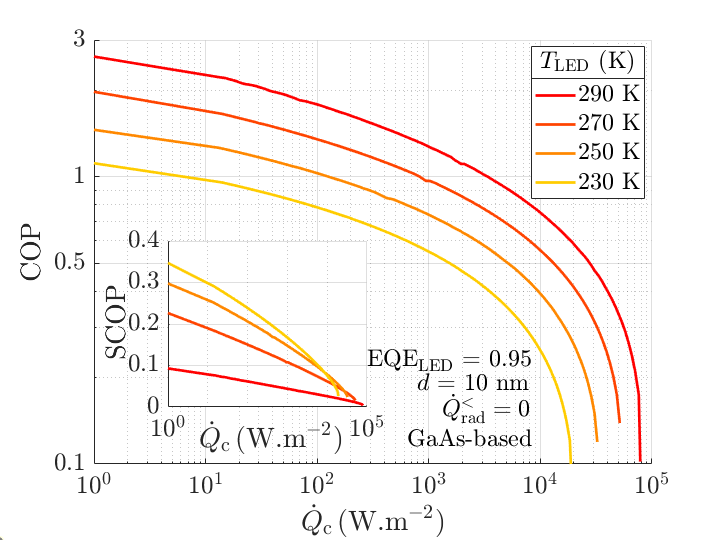}
		\label{fig:COP10nm NoSub_QE095}
	\end{subfigure}
	\begin{subfigure}{0.45\textwidth}
		\caption{}
		\includegraphics[width=\textwidth]{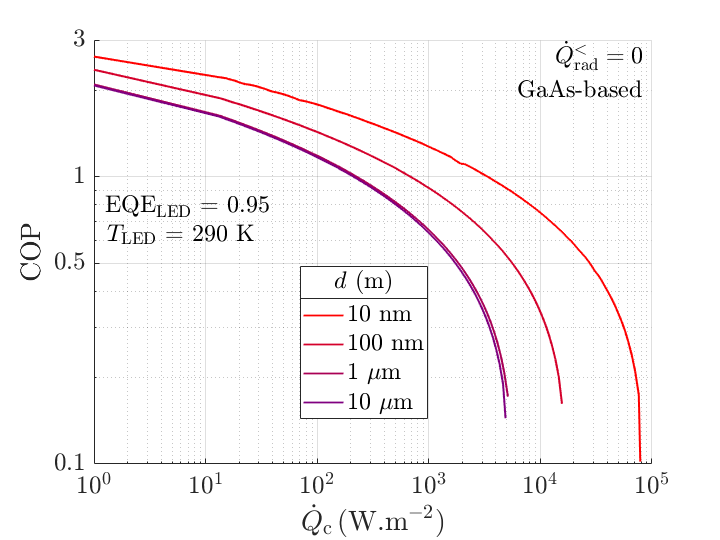}
		\label{fig:COP290K NoSub_QE095}
	\end{subfigure}
	\caption{ Performance plot of the near-field thermophotonic cooling system for \color{black}$\eqeled$\color{black} = 0.95. (a) Dependence on LED temperature including thermal radiation. (b) Dependence on distance including thermal radiation. (c) Dependence on LED temperature excluding thermal radiation. The insert provides the same data with a linear SCOP scale. (d) Dependence on distance excluding thermal radiation.}
	\label{fig:COPSCOP QE095}
\end{figure}

%Figure sur Signal au cas ou
All the previous figures are obtained in the radiative limit. In Fig. \ref{fig:COPSCOP QE095}, \color{black}$\eqeled \:$\color{black} is set to 0.95. This 5 \% drop of \color{black}$\eqeled \,$\color{black} results in a maximum COP going from $\approx$ 4.38 to 0.055 and for a cooling power decrease from 1.25 $\times$ 10$^{6}$ W.m$^{-2}$ to 2.76 $\times$ 10$^{4}$ W.m$^{-2}$ 
\color{black}
if thermal radiation is included as shown in Fig. \ref{fig:COPSCOP QE095}a. The 5\% drop of \color{black}$\eqeled \:$\color{black} leads to a 50-fold drop of cooling power and a 90-fold COP drop. Thus, the need for a high \color{black}$\eqeled \:$\color{black} for high-bandgap materials is highlighted again. In this case, reaching a significant COP requires to stay at  least at a 100-nm distance, as shown in Fig. \ref{fig:COPSCOP QE095}b. Excluding thermal radiation, performances become viable, as demonstrated by the   three-fold increase of the cooling power with respect to Fig. \ref{fig:COPSCOP QE095}a and the multiplication of the COP by three. In such case, no distance increase is required for maintaining significant performances as shown in Fig. \ref{fig:COPSCOP QE095}d. Best performances are obtained for smallest distances, but with a modest SCOP.%, similar to performances found in thermoelectric cooling devices with ZT = 2 as figure of merit \cite{Sadi2020}. 
 
\color{black}

\section{Status of near-field thermophotonics and alternative refrigeration technologies}
\begin{figure}[h]
	\centering
		\includegraphics[width=\textwidth]{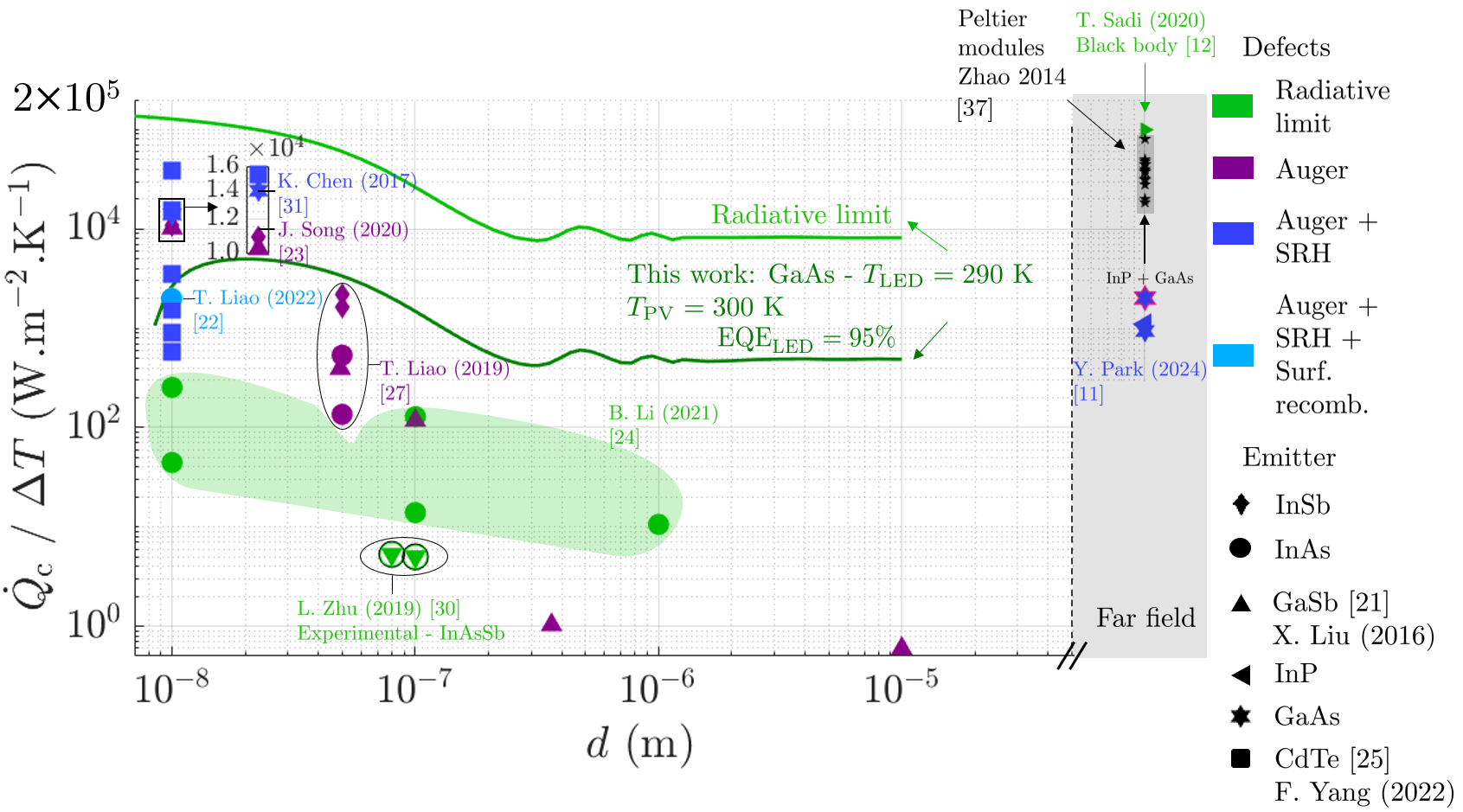}
		\caption{Cooling power potential as a function of gap distance from near-field thermophotonic data obtained so far and comparison with thermoelectrics. The solid lines represent the normalized cooling power obtained in the frame of this work for \color{black}$\eqeled$\color{black} = 1 and 0.95. %The bandgap energy for GaAs (LED) at 290 K is 1.41 eV and for AlGaAs (PV cell) at 300 K is 1.42 eV. 
For both curves, sub-bandgap radiation is included.}
		\label{fig:Biblio}
\end{figure}
\color{black}
We now compare what we have obtained with results from literature. 
Fig. \ref{fig:Biblio} shows the performances of electroluminescent cooling devices (active emitter and passive receiver) and thermophotonic cooling devices (active emitter and active receiver) as a function of gap separation between emitter and receiver. The maximum cooling power depends on the temperatures of the hot and cold sides. Therefore, to be able to compare the cooling power reported for several temperature differences, we divide it by the temperature difference, which is a commonly used metric for energy-conversion devices. % Note however that this normalization may only partly remove the impact of $\Delta T$ on the performances for thermoelectric coolers and thermophotonic harvesters with QE = 1 since their power scales as $\Delta T^2$ \cite{Legendre2024}. In the case of thermophotonic cooling, the maximum cooling power does not scale with $\Delta T^2$. 
In literature, the temperature of the receiver is most often set to 300 K. We made the same choice for consistency.  The color of each point of the figure characterizes the different nonradiative recombination mechanisms taken into account in the computations. %, with carrier concentrations taken from literature. 
The nature of the emitter and of the nonradiative recombination mechanisms can be typically linked to the bandgap energy and \color{black}$\eqeled$\color{black}, respectively. Note that as the LED bias is increased towards $\frac{E_{\text{g}}}{e}$, the most important nonradiative recombinations become Auger and surface ones \cite{Legendre2022PIN,Zarazua2016}. In the figure, the different emitter materials (see Tab. \ref{tab:bandgap_sorted}) are represented by various shapes, which link bandgap energy with associated cooling power. As can be also seen in Fig. \ref{fig:Cooling power EgQE}, emitters with low bandgap energy material such as InAs ($\Eg$ $\approx$ 0.354 eV at $T$ = 300 K) exhibit lower performances than with higher bandgap energy materials such as CdTe ($\Eg$ $\approx$ 1.45 eV at $T$ = 300 K) given the same high \color{black}$\eqeled$\color{black}. Note that InSb devices cannot be used at room temperature currently \cite{Vaillon2019} and that this datapoint is therefore only prospective.
\begin{table}[t]
	\centering
%	\rowcolors{1}{white}{gray!20}
	\begin{tabular}{c c c}
		\hline
		\textbf{Material}& & \textbf{Bandgap Energy at 300 K (eV)} \\
		\hline
		InSb & & 0.17 \\
		InAs & & 0.354 \\
		GaSb  & & 0.726 \\
		GaAs  & & 1.424 \\
		CdTe  & & 1.45 \\
		\hline
	\end{tabular}
	\caption{Bandgap energies of selected materials at 300 K. Data can be found e.g. in Ref. \cite{Kittel2005} }
	\label{tab:bandgap_sorted}
\end{table}
Several strategies have been adopted to maximize the cooling power. One of these is to excite the frustrated modes in the system. As can be found in data associated to Refs. \cite{Li2021, Liu2016}, dividing the gap size by 100 leads to a 10 to a 100-fold increase of the cooling power using the same structure. We note that almost all listed publications use mirrors behind emitters and receivers to recycle out-of-band photons. Those mirrors are expected to be as loss-free as possible. The used materials include Au \cite{Li2021}, Ag \cite{Liu2016}, Ni \cite{Song2020}, aluminium oxide Bragg reflectors \cite{PatrickXiao2018} and perfect mirrors \cite{Yang2022}. Another way of increasing the cooling power is to decrease the importance of below-bandgap radiative heat flux $\Qsub$. In deep near field, the phonon-polariton modes are prevalent and account for almost all sub-bandgap radiation for distances below $d=50$ nm. To mitigate this heat flux, Refs.  \cite{PatrickXiao2018,Song2020} use graphene on top of the LED and the PV cell, which allows tuning the surface plasmon polaritons through the chemical potential. Ref. \cite{PatrickXiao2018} shows that $\Qsub$ can be reduced up to a factor six for a 10 nm gap size.  
Our results for $\Tled$ = 290 K and $\Tpv$ = 300 K are plotted on the figure in two continuous curves. The upper curve is the idealistic upper bound (\color{black}$\eqeled$\color{black} = 1) of our system and the lower one includes some defects (\color{black}$\eqeled$\color{black} = 0.95). One can note that our results are close to those of Refs. \cite{Liao2019,Yang2022} in this nonideal case. Some authors \cite{Liao2019,Liao2022,Li2021,Zhu2019, Yang2022,Sadi2020} also report COP in their structures. Adjusting those values to SCOP, no articles provide values exceeding 35\% while including mirrors and sub-bandgap radiation. By suppressing sub-bandgap radiation but including Auger recombinations such as in Ref. \cite{Chen2015}, SCOP reaches 100 \% with powering the receiver for photon collection. To compare active radiative cooling with thermoelectric cooling (TEC) devices, data from Ref. \cite{Zhao2014} are used. More precisely, the performances of listed commercially available TEC devices are displayed in black stars, on the rightmost part of the figure. Those values are computed using the maximum cooling power divided by the surface of a given module. A one degree temperature difference is applied to obtain such high cooling power. As a result, these TEC cooling powers stand for upper bounds of the cooling power, which are on par with our own upper bound. One peculiar point in the far-field region belongs from Ref. \cite{Sadi2020}. Using ideal blackbodies and a very high bandgap of 2.5 eV at the radiative limit, the authors found an ideal cooling power of $10^{6}$ \CoolDens. 

 In general, the results from this work agree well with the data found in literature, with a cooling power per unit temperature in a range of $10^3$-$10^5$ \CoolDens K$^{-1}$. Near-field effects improve by a factor 10 the reachable cooling power. %, theoretically closing the gap with Peltier modules. 
High-bandgap energy materials such as GaAs and CdTe show improved performances compared to materials with lower bandgaps. %This is in agreement with the conclusion from Fig.\ref{fig:Cooling power EgQE}. Finally, 
Even though we considered the effect of non-radiative recombination by setting values for \color{black}$\eqeled$\color{black}, our results match well with works accounting for the recombination mechanisms more rigorously. Finally, thermophotonic devices can, in theory, reach cooling performances close to Peltier modules. There is, however, a lack of experimental results \cite{Zhu2019} confirming those tendencies. %It is to be expected that future experimental realizations, currently under development, should be able to improve over the first experimental demonstration  by \cite{Zhu2019} using biased semiconductors.  

 To finish, we note that we have addressed the case of usual refrigeration (i.e. decreasing temperature below ambient), but one can also use such a system to evacuate heat from a hot source. For instance, thermophotonic energy harvesting (hot LED and room-temperature PV cell) is being considered, where an increased dissipation rate from the hot side is possible by feeding it electrically \cite{Harder2003,Legendre2022}. Another option is to reverse-bias the hot side in what is then called a thermoradiative-PV engine. Details about various types of so-called dual radiative heat engines can be found in Ref. \cite{Legendre2025}.
\color{black}

\section{Conclusion}

By means of near-field radiation computation, cooling conditions have been established for the thermophotonic system. Using a reference material, we have come to the conclusion that a high bandgap material (above 1 eV) and large \color{black}$\eqeled$\color{black} ($>$ 0.95) allows for a cooling power larger than $10^{3}$ \CoolDens, which is necessary to compete with thermoelectrics. Low \color{black}$\eqeled$\color{black} prevent from obtaining significant cooling power. The computations of a GaAs-based TPX system result in a cooling power that can reach 1.46 $ \times$ 10$^{6} $ W.m$^{-2}$. We have underlined that the increase of the cooling power, expected when the distance becomes smaller, is limited down to a certain gap size. Indeed, thermal radiation emitted by the PV cell, which acts as a heat leak in the system, becomes particularly detrimental for small gap sizes, making a certain distance optimal to maximise $\Qc$. Finally, we have compared our results with the existing literature on the subject and found that near-field thermophotonic refrigerators could indeed compete with thermoelectric coolers: our upper bound matches that of TEC cooling devices. As thermophotonic systems are still under development, and our system is not thoroughly optimized, better cooling performances than those found here could be achieved. In particular, it was found for energy-harvesting thermophotonic devices that solving the electrical transport by means of drift-diffusion equations \cite{Legendre2022PIN} can outperform solutions from the detailed-balance approach. Structuring the materials as multilayers could also increase the performances, by filtering the sub-bandgap energy radiative transfer and reducing the spectral width above bandgap. One should note that including resistive losses should however be taken into account and could reduce the performances.

\footnotesize

Acknowledgements: 
We thank P. Kivisaari, J. Van Gastel, M. Thomas, W. Sghaier, K. Tappura for the constructive discussions. We thank also anonymous reviewers for enlightening remarks. This work has received funding from European Union Horizon 2020 research program through EU projects OPTAGON (GA 964698) and TPX-Power (GA 951976).

\bibliography{Performancesoffarandnear-fieldthermophotonicrefrigeration}% Produces the bibliography via BibTeX.

\end{document}